\documentclass[10pt, conference, letterpaper]{IEEEtran}
\usepackage{url}
\IEEEoverridecommandlockouts
\usepackage{cite}
\usepackage{amsmath,amssymb,amsfonts}
\usepackage{algorithmic}
\usepackage{graphicx}
\usepackage{textcomp}
\usepackage{xcolor}
\def\BibTeX{{\rm B\kern-.05em{\sc i\kern-.025em b}\kern-.08em
    T\kern-.1667em\lower.7ex\hbox{E}\kern-.125emX}}
\usepackage{xspace}
\usepackage{multirow}
\usepackage{epsfig, subfigure, wrapfig, amsmath, amssymb}
\usepackage{enumitem,kantlipsum}

\definecolor{red}{rgb}{1,0,0}

\begin{document}

\title{BIoTA: Control-Aware Attack Analytics for Building Internet of Things}

\author{
\IEEEauthorblockN{Nur Imtiazul Haque\IEEEauthorrefmark{1}, Mohammad Ashiqur Rahman\IEEEauthorrefmark{1}, Dong Chen\IEEEauthorrefmark{2}, and Hisham Kholidy\IEEEauthorrefmark{3}}
\IEEEauthorblockA{\IEEEauthorrefmark{1}Analytics for Cyber Defense (ACyD) Lab, Florida International University, USA\\
	\IEEEauthorrefmark{2}Cyber-Physical Systems Laboratory (CPSLab), Florida International University, USA\\
	\IEEEauthorrefmark{3}Department of Network and Computer Security, State University of New York Polytechnic Institute, USA\\
	\IEEEauthorrefmark{1}\{nhaqu004, marahman\}@fiu.edu, \IEEEauthorrefmark{2}dochen@cis.fiu.edu, \IEEEauthorrefmark{3}kholidh@sunypoly.edu
	}
}


\thispagestyle{plain}
\pagestyle{plain}

\maketitle

\begin{abstract}
Modern building control systems adopt demand control heating, ventilation, and cooling (HVAC) for increased energy efficiency. The integration of the Internet of Things (IoT) in the building control system can determine real-time demand, which has made the buildings smarter, reliable, and efficient. As occupants in a building are the main source of continuous heat and $CO_2$ generation, estimating the accurate number of people in real-time using building IoT (BIoT) system facilities is essential for optimal energy consumption and occupants' comfort. However, the incorporation of less secured IoT sensor nodes and open communication network in the building control system eventually increases the number of vulnerable points to be compromised. Exploiting these vulnerabilities, attackers can manipulate the controller with false sensor measurements and disrupt the system's consistency. The attackers with the knowledge of overall system topology and control logics can launch attacks without alarming the system. This paper proposes a building internet of things analyzer (BIoTA) framework\footnote{https://github.com/imtiazulhaque/research-implementations/tree/main/biota} that assesses the smart building HVAC control system's security using formal attack modeling. We evaluate the proposed attack analyzer's effectiveness on the commercial occupancy dataset (COD) and the KTH live-in lab dataset. To the best of our knowledge, this is the first research attempt to formally model a BIoT-based HVAC control system and perform an attack analysis.
\end{abstract}


\section{Introduction}
\label{sec:introduction}
The heating, ventilation, and cooling (HVAC) control system accounts for almost 35\% of energy consumption in residential and commercial building facilities~\cite{encycle2020}. With the advent of the building internet of things (BIoT), HVAC systems' energy consumption is reducing significantly~\cite{bajer2018iot}. Estimating the accurate number of occupancy through sensor nodes, BIoT can make optimal control decisions that benefit modern HVAC control systems to maintain precise mechanical ventilation and cooling~\cite{akkaya2015iot}. In a BIoT-based HVAC control system, occupancy, temperature, and $CO_2$ monitoring sensor nodes send periodic measurements to the controller. The controller accumulates all the sensor status and generates control decisions accordingly to regulate actuators (e.g., air mixing vents and supply fans). The integration of IoT in the HVAC control system adds several benefits concerning monitoring, analyzing, and tracking system status and generating automated control decisions. Additionally, if some inconsistent events occur, real-time alerts are sent to the system admin to take necessary actions.

Current building architecture implants distributed air handling units (AHUs) throughout the building for maximum energy efficiency~\cite{yu2010integrating}. Each AHU is responsible for maintaining the indoor air quality (IAQ) and the temperature of a specific zone using the outside fresh air, and the recirculating return air from the zone. Several factors contribute to energy consumption by an HVAC control system (e.g., weather condition, peak demand time, occupancy pattern, etc.). At the same time, occupants' comfort depends on IAQ and the desired set-point temperature. Allowing fresh outdoor air in the zones helps control the air quality, but the desired room temperature deviates significantly from the fresh air temperature. 
The HVAC control systems are prone to utilize recirculating the zone air for energy efficiency, which, in turn, raises the concentration of air pollutants as occupants of the zones always produce $CO_2$. Therefore, for optimal control strategy, amalgamated fresh and return air is supplied through the zones to confirm comfort and cost-efficiency.

However, BIoT-enabled HVAC control systems are highly potential for cyberattacks. 
According to a report by Qualys, a cloud security service provider connecting almost 55,000 HVAC systems to the internet possess flaws that can easily be exploited by hackers~\cite{target2014}.
Cybersecurity firm ForeScout Technologies reported in 2019 that 8,000 IoT devices related to HVAC systems are vulnerable to cyberattacks~\cite{asmag2019}. 
In Finland, an attack on the heating system left residents of two apartment buildings in the cold for nearly a week~\cite{IBTimes}. An adversary, depending on the attack capability, can achieve various attack goals exploiting these vulnerabilities. Therefore, it is essential to explore these wide-ranged potential attacks on a smart HVAC system considering various attack capabilities and thus understand the system's attack-resiliency in different scenarios. Such an analysis cannot be done empirically through experimentations; instead, it requires non-invasive attack analytics.

This paper addresses the challenge mentioned above by proposing the BIoT analyzer (BIoTA), a formal framework
to perform security analysis of smart building HVAC control systems. In particular, this formal analysis considers two distinct kinds of attack targets: (i) increasing energy consumption and (ii) disrupting occupants' comfort for maximum occupants. We think of a BIoT-enabled smart building, where the controller also looks for suspected behaviors (e.g., zone occupancy exceeding maximum capacity, inconsistency in sensor measurements, etc.) along with generating control decisions. Hence, an attacker cannot launch an attack manipulating sensor measurements arbitrarily. Our proposed framework synthesizes feasible attacks based on the attacker's knowledge, accessibility, and target/goal, which allows us to realize the system's attack-resiliency according to the attack model. 
In this work, we are formally modeling the HVAC control system leveraging mass balance equation-based ventilation control and energy balance equation-based temperature control, which can imitate the dynamics of air and heat balance of a HVAC control system~\cite{lu2011novel}. Moreover, we analyze both commercial and residential buildings' security using two state-of-the-art building occupancy datasets naming commercial occupancy dataset (COD)~\cite{liu2017cod} and KTH live-in lab dataset~\cite{kth2020}.

The rest of the paper is organized as follows: We briefly discuss HVAC control systems in Section~\ref{sec:background}. We present the BIoTA framework's architecture in Section~\ref{sec:biota-framework}. In Section~\ref{sec:formal-analysis-model}, we present the formal model of attack analysis.  We provide two case studies in Section~\ref{sec:case-study}.  We present the evaluation results in Section~\ref{sec:evaluation}. Finally, we conclude the paper in Section~\ref{sec:conclusion}.


\section{Background}
\label{sec:background}
In this section, we briefly discuss the HVAC control system's working principle and the attack model.

\begin{figure}[!t]
    \centering
    \includegraphics[width=0.9\columnwidth]{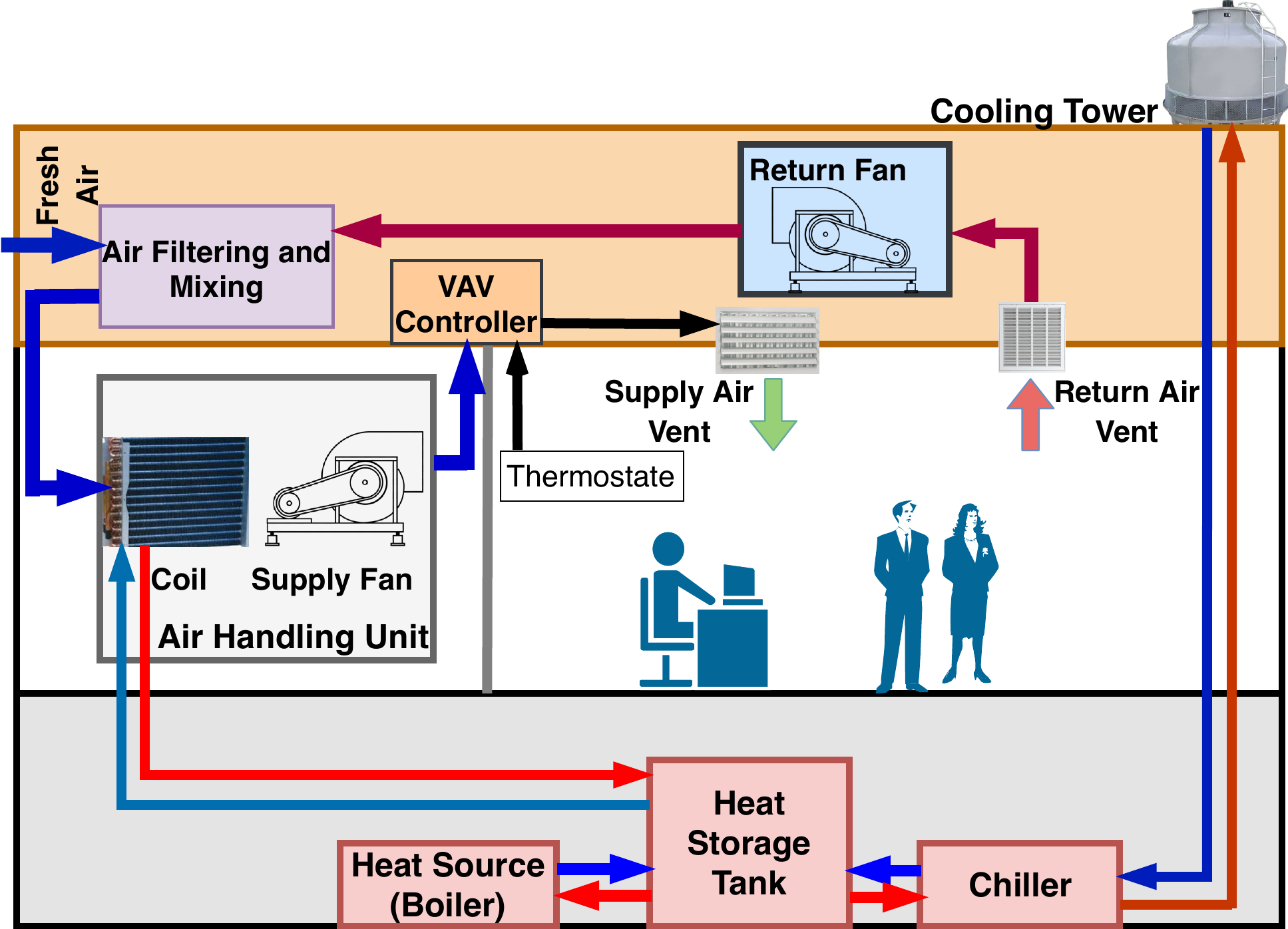}
    \vspace{-6pt}
    \caption{HVAC Control Setup in Smart Building}
    \vspace{-18pt}
    \label{fig:hvac_control}
\end{figure}

\vspace{-3pt}
\subsection{HVAC Control System}
\vspace{-3pt}
A simplistic view of a building HVAC control system's overall control architecture is demonstrated in Fig.~\ref{fig:hvac_control}. The air handling unit (AHU) is the heart of HVAC control, which performs mixed air filtering, humidity control through humidifier and dehumidifier, as well as air temperature control using heating and cooling coils~\cite{leong2019fault}. After proper air quality control, the AHU supplies air to the zones using the supply fan. The operation of the supply fan is scheduled by the control system, depending on the zone demand. The AHU brings air from a duct, where both fresh outside air and recirculated return air from the rooms are mixed and filtered. A zone's demand is determined by a thermostat, which is used to set the preferred temperature, humidity, and $CO_2$ concentration of the zone occupants. 

Each room of a zone has different air demand because of the difference in corresponding heating or cooling loads. The cooling and heating loads determine the amount of heat energy that needs to be removed or added in the building zone. A valve is placed in the vents to control the dampers' opening and maintain the required airflow. The temperature of the zone's supply air depends on the temperature of the mixed air and zone temperature setpoints. In the case of cooling, the temperature of supply air is typically set to 55\textdegree F, while for heating, supply air temperature is kept 40-70\textdegree F above the return air temperature~\cite{wang2012air,inspectapedia2020}. The air is chilled or heated through the refrigerants flowing through the coils. After heating or chilling air, the refrigerant temperature gets reduced, which is brought back to the coil setpoint temperature using the heat storage tank typically placed in the basement of the building. The heat storage tank's temperature is maintained using the connected chiller and the boiler based on the weather demand. The chillers are connected to the rooftop cooling tower, which brings air and water together into direct contact for reducing refrigerants' temperature.

\vspace{-3pt}
\subsection{Ventilation and Temperature Control}\vspace{-3pt}
The ventilation control is required to maintain the indoor air quality (IAQ). The IAQ can be considered satisfactory, provided the $CO_2$ concentration of the occupied zone is less than 1000 parts per million (ppm)~\cite{Ohsonline2016}. The $CO_2$ concentration of the fresh outside air stays near about 400 ppm. The steady-state ventilation control of smart building HVAC control system follows the dynamics of the mass balance equation to ensure derivative of $CO_2$ concentration over the time is equal to zero and thus maintains the IAQ to a specific setpoint~\cite{cali2015co2}. On the other hand, temperature control obeys the energy balance equation for dealing with the cooling and heating loads of the room and keeps the room temperature steady to a comfortable temperature. The HVAC control system's main challenge for temperature and ventilation control arises from the sudden change of demand because of the varying occupancy rate, generating heat, and $CO_2$. The ventilation control tries to utilize more fresh outside air for a faster reduction of the $CO_2$ concentration from the air. But using more fresh air increases control cost due to significant deviation between fresh air and supply air temperature.  The return air temperature stays closer to the zone's setpoint temperature and supply air temperature. However, the return air is contaminated with zone pollutants, which makes $CO_2$ concentration in the return air to be high. 

\vspace{-3pt}
\subsection{HVAC Cost Calculation}\vspace{-3pt}

Hence, in order to satisfying the occupants' need for ventilation and temperature comfort, both fresh air from outside needs to be brought in and return air from the zones needs to be re-circulated. The effectiveness of the HVAC control system depends on optimally mixing fresh and return air to satisfy both energy consumption and occupants' comfort. After proper mixing, the HVAC control system measures the temperature and $CO_2$ concentration of the mixed air. If the mixed air temperature is high, the cool refrigerant flowing through the coil chills down the hot air to the supply air temperature. As the air is coming in very warm and going out very cool, the air gets condensed in the form of water, and that water is removed from the system using a pipe. The mass flow rate of condensate water and mixed air are used with few other psychrometric parameters to calculate the energy used in the coil~\cite{dyro2004clinical}. During this process of cooling, the refrigerant temperature gets high, which requires bringing back to normal temperature in the chiller. The mixed air processing and cooling refrigerants in the chiller are mainly responsible for the HVAC cooling control cost. The air heating follows a similar procedure.

\vspace{-3pt}
\subsection{BIoT Architecture of HVAC Control System}\vspace{-3pt}
For efficient control, information of the zones must be shared, which creates the need for the internet of things-based models. In our work, we consider a BIoT system, where occupancy, temperature, and $CO_2$ sensor nodes send their measurements to a central controller for optimal control decisions. The controller generates control signals to actuate the supply fan, return fan, and the valves of the vents. 
%
Our HVAC modeling assumes the followings:
\begin{enumerate}[wide, labelwidth=!, labelindent=0pt]
    \item The cooling load and ventilation requirement of every room of a particular zone are similar.
    \item The cooling load of summer season and heating load of the winter season are presumed to be equal and constant throughout the year. 
    \item The occupants of a particular zone radiates an equal amount of heat and exhales an equal amount of $CO_2$ in every time instance. The occupants generated heat and $CO2$ amount in a particular zone depends on the level on average occupant's physical activity, age, and metabolism rate~\cite{persily2017carbon}.  
    \item The sensors have no sensitivity issues and precision errors.
\end{enumerate}

\subsection{BIoT HVAC Control Attack Model}
The BIoT architecture opens a massive cyberattack surface for the attackers. By compromising sensor nodes and the communication link between sensors and controller, an attacker can inject false sensor measurements and propel the controller to make wrong decisions. Most recent research has shown that HVAC sensors can be compromised. For instance, Tu et al. conducted a physical level attack on the temperature sensor of an infant incubator which also prevents triggering automatic temperature alarm~\cite{tu2019trick}. In addition, leveraging a man-in-the-middle (MitM) attack in the open communication channel of a BIoT-based model, an attacker can sniff and tamper sensor measurements. Newaz et al. demonstrated that adversarial machine learning based attack generation technique can be adopted for compromising sensor measurements without alarming the system~\cite{newaz2020adversarial}. As the sensor measurements of the BIoT system can be altered due to adversarial intent or sensor faults, the controller undertakes several security measures.  To model attacks, we consider that the incoming and outgoing events of the buildings are securely monitored through biometric sensors. Thus, malicious intent to increase the number of occupants in the targeted zones will notify the controller. Moreover, the maximum capacity for occupants in each zone is verified at every time slot by the controller. 

\vspace{-3pt}
\subsection{Related Work}
\label{subsec:related-works}\vspace{-3pt}
Research on smart building security recently has received special focus. Mace et al. introduced a multi-model methodology for assessing the security of smart-building systems, utilizing a suite of modeling, simulation, and analysis tools~\cite{mace2018multi}. Meyer et al. presented an abstract model of a building automation system using attack trees in order to simplify the threat identification model~\cite{meyer2016threat}. In particular, the attack trees were used to analyze threats where an attacker could get sensor information and control actuators. Mustafa et al. analyzed the security of a smart building or building automation system using attack tree analysis tools based on a high-level requirement of safety and security~\cite{mustafa2016ata}. Abdulmunem et al. also employed attack tree analysis and Markov models to analyze a building automation system's security and availability during the building's life cycle~\cite{abdulmunem2016availability}. 

Model-driven security analysis-based researches of smart building control are also gaining popularity. Hachem et al. proposed a model-driven engineering method, systems-of-systems security (SoSSec), which comprises a modeling language (SoSSecML) to model system-of-systems (SoS) and multi-agent systems for security analysis of smart building's SoS architectures~\cite{hachem2020modeling}. Moreover, Haque et al. has shown formal modeling based attack analytics in a machine learning-based smart healthcare system~\cite{haque2021novel}.

However, none of these works can perform automated attack analysis of a BIoT-based HVAC control system based on an attack model, a given set of adversarial capabilities, or an attack impact. Our proposed formal framework synthesizes formally proven attack vectors that can achieve an attack goal.

\section{BIoTA Framework}
\label{sec:biota-framework}

\begin{figure}[!t]
    \centering
    \includegraphics[width=1\columnwidth]{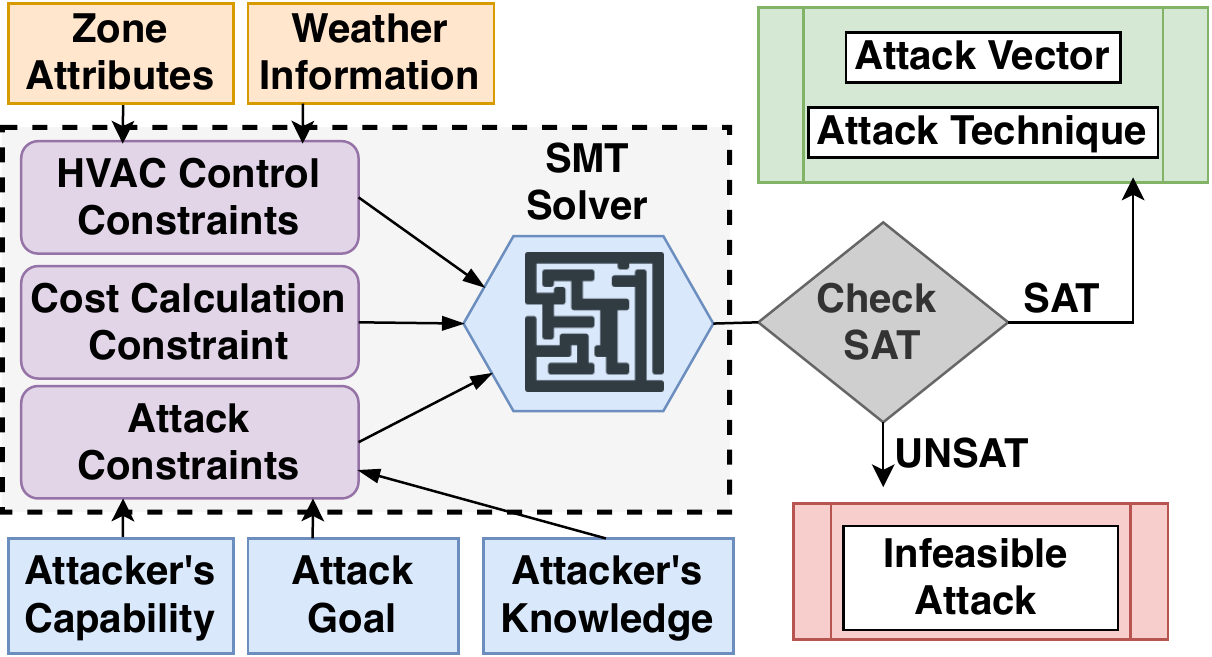}
    \vspace{-15pt}
    \caption{BIoTA Framework}
    \vspace{-9pt}
    \label{fig:biota_framework}
\end{figure}

The architecture of the proposed framework is shown in Fig.~\ref{fig:biota_framework}. In this framework, the HVAC control constraints are generated using the building floor plan, properties of the zones, and the current states of sensor measurements. The attack constraints are dependent on the attacker's knowledge, accessibility to sensor measurements, and attack goal. A set of constraints related to HVAC cost calculation are also needed to determine the cost of control before and after an attack to check whether the attack goal is satisfied or not. All these constraints are passed to the satisfiability modulo theory (SMT)-based solver as a constraint satisfaction problem (CSP). The solver solves the CSP utilizing various effective theories. If there is any possible solution, the SMT solver is guaranteed to find that which is indicated by SAT result. An UNSAT result, on the other hand, signifies that based on the current attacker's accessibility, the attack goal cannot be attained. When the solver approves the attack's feasibility, it also outputs the possible way of attack showing the attack vector. The attack vector discloses the complete attack technique showing which sensor measurements need to the altered along with the measurement values.  

\section{Formal Modeling of 
Attack Analysis}
\label{sec:formal-analysis-model}
In this section, we discuss the formal model of attack vector synthesis in detail.
\begin{table}[!t]
\centering
\caption{Modeling Notations} 
\label{tab:modeling-notations}
\vspace{-6pt}
\scriptsize
\begin{tabular}{|c|p{4.6cm}|p{1.2cm}|p{0.5cm}|}
\hline
\textbf{Notation} & \textbf{Description} & \textbf{Unit} & \textbf{Data Type} \\ \hline
$P$ & Atmospheric pressure & Pa & Real \\ \hline
${V^{Z}_j}$            & Volume of j-th zone    & $Ft^3$    & Real                \\ \hline
${C^{P}_{j}}$            & $CO_2$ emission per person per minute at j-th zone &  $CFM$   & Real                \\ \hline
${Q^{P}_{j}}$            & Heat radiation per person per second at j-th zone &  kW   & Real                \\ \hline
${Q^{L}_{ij}}$            & Cooling or heating load of j-th zone at i-th time instance &  kW   & Real                \\ \hline

${C^{set}_{ij}}$            & $CO_2$ setpoint of j-th zone at i-th time instance &  PPM   & Integer                \\ \hline
${T^{set}_{ij}}$            & Temperature setpoint of j-th zone at i-th time instance &  \textdegree F   & Integer                \\ \hline

${\mathcal{S}^{to}_{i}}$            & Temperature of fresh outdoor air at i-th time instance & \textdegree F    & Real                \\ \hline
${\mathcal{S}^{co}_{i}}$            & $CO_2$ concentration of fresh outdoor air at i-th time instance & PPM    & Real                \\ \hline

${\mathcal{S}^{ti}_{ij}}$             & indoor temperature of j-th zone at i-th time instance & PPM    & Real                \\ \hline
${\mathcal{S}^{ci}_{ij}}$             & indoor $CO_2$ concentration of j-th zone at i-th time instance & PPM    & Real                \\ \hline
${\mathcal{S}^{tm}_{ij}}$            & Temperature of mixed air of j-th zone at i-th time instance &  \textdegree F   & Real                \\ \hline

${\mathcal{S}^{cm}_{ij}}$            & $CO_2$ concentration of mixed air at j-th zone at i-th time instance &  PPM   & Real                \\ \hline
${\mathcal{S}^{hm}_{ij}}$            & Relative Humidity of mixed air of j-th zone at i-th time instance &  (\%)   & Real                \\ \hline
${\mathcal{S}^{ts}_{ij}}$            & Temperature of supply air of j-th zone at i-th time instance & \textdegree F    & Real                \\ \hline
${\mathcal{S}^{hs}_{ij}}$            & Relative humidity of supply air of j-th zone at i-th time instance & (\%)    & Real                \\ \hline

${\mathcal{S}^{o}_{ij}}$            & Number of occupants j-th zone at i-th time instance &  Person   & Integer                \\ \hline

${\dot{v}^{m}_{ij}}$            & Volumetric flow of mixed air of j-th zone at i-th time instance &  CFM   & Real                \\ \hline

${\dot{m}^{m}_{ij}}$            & Mass flow of mixed air j-th zone at i-th time instance &  $kgs^{-1}$   & Real                \\ \hline

$\Delta t$            & Sampling time of the controller &  minute   & Integer                \\\hline

${Pw^{m}_{ij}}$            & Mixed air partial pressure of water vapor of j-th zone at i-th time instance &  Pa   & Real                \\ \hline

${Pw^{s}_{ij}}$            & Supply air partial pressure of water vapor of j-th zone at i-th time instance &  Pa   & Real                \\ \hline

${Sh^{m}_{ij}}$            & Specific heat of mixed air of j-th zone at i-th time instance &  Jkg$^{-1}$K$^{-1}$   & Real                \\ \hline

${Sh^{s}_{ij}}$            & Specific heat of supply air of j-th zone at i-th time instance &  J$kg^{-1}K^{-1}$   & Real                \\ \hline
${Sh^{W}}$            & Specific heat of water &  Jkg$^{-1}K^{-1}$   & Real                \\ \hline

${\dot{m}^{cl}_{j}}$            & Mass flow of air in the coil in j-th zone &  $kgs^{-1}$   & Real                \\ \hline
${Eth^{m}_{ij}}$            & Enthalpy of mixed air of j-th zone at i-th time instance & Jkg$^{-1}$    & Real                \\ \hline
${Eth^{s}_{ij}}$            & Enthalpy of supply air of j-th zone at i-th time instance & J$kg^{-1}$    & Real                \\ \hline

${T^{cnd}_{ij}}$            & Temperature of condenser water of j-th zone at i-th time instance & \textdegree F    & Real                \\ \hline
${T^{cl}_{ij}}$            & Temperature of coil water of j-th zone at i-th time instance & \textdegree F    & Real                \\ \hline

$Cl^{set}$ & Temperature setpoint of coil refrigerant & \textdegree F & Real     \\ \hline
${\dot{m}^{cnd}_{ij}}$            & Mass flow of condenser water of j-th zone at i-th time instance &  $kgs^{-1}$   & Real                \\ \hline

$\mathcal{S_j}$            & Set of all Sensors of j-th zone &  PPM, \textdegree F, Person   & Set                \\ \hline

$\mathcal{T}$            & Set of all timeslots &  Unit   & Set               \\ \hline
$\mathcal{Z}$            & Set of all Zones &  Unit   & Set                \\ \hline
$\mathcal{S}$            & Set of all Sensors &  PPM, \textdegree F, Person   & Set                \\ \hline
\end{tabular}
\vspace{-9pt}
\normalsize
\end{table}

\subsection{Formal Modeling of HVAC Control System}
For analyzing the smart building HVAC control system attacks, BIoTA formally models the control system using mathematical equations considering underlying system dynamics.
Table~\ref{tab:modeling-notations} demonstrates the modeling notations for HVAC control. We consider the mass balance equation for ventilation control. Cali et al. approximated a numerical solution for the mass balance equation that can be utilized for taking ventilation control decisions \cite{cali2015co2}. The effectiveness of the controller depends on the accurate estimation of the number of people at a particular zone. We consider Passive Infra-red (PIR) sensors to measure the number of persons at a particular time instance. The ventilation requirement of a particular zone can be calculated using the following formula.
%
\begin{equation}
    \label{eq:ventilation_control}
    \begin{split}
    \forall_{i \in \mathcal{T}, j \in \mathcal{Z}}\frac{\mathcal{S}^{o}_{ij} C^{P}_{i}\Delta t}{V^{Z}_i} = {C^{set}_{ij}} & - \left(1-\frac{{\dot{v}^{m}_{i}}\Delta t}{V^{Z}_i}\right)\mathcal{S}^{ci}_{ij} \\
    & - \frac{ \dot{v}^{m}_{i} \Delta t}{V^{Z}_i}\mathcal{S}^{cm}_{ij}
    \end{split}
\end{equation}
%
Here, $\dot{v}^{m}_{i}$ is the mixed air volumetric flow which is the combination of both fresh and return air. Equation~\ref{eq:ventilation_control} is used to calculate the required amount of ventilation at each time instances. The temperature control decision is made based on the following equation.
\begin{equation}
    \label{eq:temperature_control}
    \begin{split}
    \forall_{i \in \mathcal{T}, j \in \mathcal{Z}}{\dot{m}^{m}_{ij} Sh^{m}_{ij} (T^{set}_{ij} - \mathcal{S}^{ts}_{ij})}= Q^{L}_{j} +
    \mathcal{S}^{o}_{ij} Q^{P}_{j}
    \end{split}
\end{equation}

The controller also verifies the consistency of sensor measurements of the last control decision. 
\begin{equation}
    \label{eq:ventilation_control_ver}
    \begin{split}
    \forall_{i \in \mathcal{T}, j \in \mathcal{Z}}\frac{\mathcal{S}^{o}_{(i-1)j} C^{P}_{i}\Delta t}{V^{Z}_i} & =    \mathcal{S}^{ci}_{ij}  - \\ \left(1-\frac{{\dot{v}^{m}_{i}}\Delta t}{V^{Z}_i}\right) & \mathcal{S}^{ci}_{(i-1)j} - \frac{ \dot{v}^{m}_{i} \Delta t}{V^{Z}_i}\mathcal{S}^{cm}_{ij}
    \end{split}
\end{equation}
%
\begin{equation}
    \label{eq:temperature_control_ver}
    \begin{split}
    \forall_{i \in \mathcal{T}, j \in \mathcal{Z}}{\dot{m}^{m}_{ij} Sh^{m}_{ij} (\mathcal{S}^{ti}_{(i-1)j} - \mathcal{S}^{ts}_{ij})}= Q^{L}_{j} +
    \mathcal{S}^{o}_{(i-1)j} Q^{P}_{j}
    \end{split}
\end{equation}

The temperature of supply air is kept 55 \textdegree F in summer season and 120 \textdegree F for winters. For effective control, the temperature of supply air is slightly increased at the time of cooling, while it is slightly reduced for heating.
\begin{equation}
    \label{eq:temperature_control}
    \begin{split}
    \forall_{i \in \mathcal{T}, j \in \mathcal{Z}} 55 \leq \mathcal{S}^{ts}_{ij} \leq 120
    \end{split}
\end{equation}

\subsection{Formal Modeling of HVAC Cost Calculation}
BIoTA framework considers the cooling cost of a zone, which depends on the supply and the mixed air temperature. For chilling the mixed air to supply air, a significant amount of energy is needed in the cooling coil, which in terms increases the temperature of the refrigerants of the cooling coil. The refrigerants are sent to the chiller tower for cooling to the setpoint temperature, which also requires similar energy needed at the cooling coil. 

For determining the psychrometric values, six co-efficient values are needed according to ASHRAE standard ($cf_0$= -5800.22, $cf_1$= 1.39, $cf_2$ = -0.049, $cf_3$ = 4.17 $\times 10^{-5}$, $cf_4$ = -1.44 $\times 10^{-8}$, $cf_5$ = 6.54). Calculating partial pressure of water is needed to determine specific heat, specific volume, and enthalpy of the supply and mixed air. 
\begin{equation}
\begin{split}
    \forall_{i \in \mathcal{T}, j \in \mathcal{Z}}{Pw^{m}_{ij}} = e^{\sum_{l=0}^{4}{(cf_{l}\mathcal{S}^{tm}_{ij}})^(l-1) + cf_5log_e(\mathcal{H}^{k}_{ij})} \mathcal{S}^{hm}_{ij}
\end{split}
\end{equation}
\begin{equation}
\begin{split}
    \forall_{i \in \mathcal{T}, j \in \mathcal{Z}}{Pw^{s}_{ij}} = e^{\sum_{l=0}^{4}{(cf_{l}\mathcal{S}^{ts}_{ij}})^(l-1) + cf_5log_e(\mathcal{H}^{k}_{ij})} \mathcal{S}^{hs}_{ij}
\end{split}
\end{equation}

The specific heat computation is needed to determine the mass airflow rate of condenser water, which in turn, is necessary for calculating the cost associated with coil cooling/heating.
\begin{equation}
\begin{split}
    \forall_{i \in \mathcal{T}, j \in \mathcal{Z}, k \in [m, s]}Sh^{k}_{ij} = \frac{0.621945Pw^{k}_{ij}}{P-Pw^{k}_{ij}}
\end{split}
\end{equation}

The enthalpy of the mixed and supply air is directly used to determine the HVAC cooling/heating coil cost.
\begin{equation}
\begin{split}
    \forall_{i \in \mathcal{T}, j \in \mathcal{Z}}Eth^{m}_{ij} = 1.006 * \mathcal{S}^{tm}_{ij} + Sh^{m}_{ij} * (2501 + 1.86* \mathcal{S}^{tm}_{ij})
\end{split}
\end{equation}
\begin{equation}
\begin{split}
    \forall_{i \in \mathcal{T}, j \in \mathcal{Z}}Eth^{s}_{ij} = 1.006 * \mathcal{S}^{ts}_{ij} + Sh^{s}_{ij} * (2501 + 1.86* \mathcal{S}^{ts}_{ij})
\end{split}
\end{equation}

The specific volume of mixed air is used to calculate the volumetric flow of air simply by dividing mass flow rate by specific volume.
\begin{equation}
\begin{split}
    \forall_{i \in \mathcal{T}, j \in \mathcal{Z}}Sv^{m}_{ij} = 287.042 \mathcal{S}^m_{ij}\frac{1+1.607858Sh^m_{ij}}{P}
\end{split}
\end{equation}

The mass flow of air in the condenser can be derived from the specific heat of mixed and supply air and mixed air flow rate. 
\begin{equation}
\begin{split}
    \forall_{i \in \mathcal{T}, j \in \mathcal{Z}}{\dot{m}^{cnd}_{ij}} = \dot{m}^{m}_{ij} (Sh^{m}_{ij} - {Sh}^{s}_{ij})
\end{split}
\end{equation}

The mass flow rate of condenser air can determine the heat energy required to chill the hot  mixed air to supply air setpoint temperature.
\begin{equation}
\begin{split}
    \forall_{i \in \mathcal{T}, j \in \mathcal{Z}}cost^{cl}_{ij} =  \dot{m}^{m}_{ij} * (Eth^{s}_{ij} - Eth^{m}_{ij}) + {\dot{m}^{cnd}_{i}} {Eth}^{cnd}_{ij}
\end{split}
\end{equation}

But for chilling the mixed air, the temperature of coil refrigerant rises above the required temperature, which can be determined by following equation. 
\begin{equation}
\begin{split}
    \forall_{i \in \mathcal{T}, j \in \mathcal{Z}} T^{cl}_{ij} = Cl^{set} + (cost^{cl}_{ij} / (\dot{m}^{cl}_i * Sh^{W}) ) 
    \end{split}
\end{equation}

\begin{table*}[]
\centering
\caption{Zone properties for COD and KTH datasets}
\vspace{-5pt}
\scriptsize
\label{tab:testbed-description}
\begin{tabular}{|l|l|c|c|c|c|c|}
\hline
\textbf{Datasets}              & \textbf{Zones}       & \multicolumn{1}{l|}{\textbf{Volume ($ft^3$)}} & \multicolumn{1}{l|}{\textbf{\begin{tabular}[c]{@{}l@{}}CO2 Emission/Person (cfm)\end{tabular}}} & \multicolumn{1}{l|}{\textbf{\begin{tabular}[c]{@{}l@{}}Heat Radiation/Person (kW)\end{tabular}}} & \multicolumn{1}{l|}{\textbf{\begin{tabular}[c]{@{}l@{}}Cooling Load (kW)\end{tabular}}} & \multicolumn{1}{l|}{\textbf{\begin{tabular}[c]{@{}l@{}}Capacity (Person)\end{tabular}}} \\ \hline
\multirow{4}{*}{\textbf{COD}} & \textbf{Entrance}    & 12570.36                                                     & 0.0221                                                                                                 & 0.1082                                                                                                & 0.725                                                                                     & 50                                                                                                          \\ \cline{2-7} 
                              & \textbf{Clemente}    & 11687.61                                                     & 0.0213                                                                                                 & 0.1071                                                                                                & 0.300                                                                                     & 10                                                                                                          \\ \cline{2-7} 
                              & \textbf{Warhol}      & 10910.79                                                     & 0.0183                                                                                                 & 0.0924                                                                                                & 0.450                                                                                     & 25                                                                                                          \\ \cline{2-7} 
                              & \textbf{Laboratory}  & 17937.48                                                     & 0.0254                                                                                                 & 0.1193                                                                                                & 0.825                                                                                     & 40                                                                                                          \\ \hline
\multirow{3}{*}{\textbf{KTH}} & \textbf{Living Room} & 1119.2                                                       & 0.0183                                                                                                 & 0.0924                                                                                                & 0.05                                                                                      & 5                                                                                                           \\ \cline{2-7} 
                              & \textbf{Kitchen}     & 344.8                                                        & 0.0221                                                                                                 & 0.1082                                                                                                & 0.02                                                                                      & 2                                                                                                           \\ \cline{2-7} 
                              & \textbf{Bathroom}    & 252.2                                                        & 0.0254                                                                                                 & 0.1193                                                                                                & 0.02                                                                                      & 1                                                                                                           \\ \hline
\end{tabular}
\normalsize
\vspace{-9pt}
\end{table*}
Hence, the refrigerants are passed to the chiller for cooling it back to the normal temperature which also adds cost to the system. The cost can be estimated by the following equation.
\begin{equation}
\begin{split}
    \forall_{i \in \mathcal{T}, j \in \mathcal{Z}}cost^{chil}_{ij} = \dot{m}^{cl}_i * Sh^{W} * (T^{cl}_{ij} - Cl^{set})
\end{split}
\end{equation}

The overall cost of HVAC control cost is dependent on the coil cost and the chiller cost.
\begin{equation}
\begin{split}
    TC^{HVAC} = \sum_{i \in \mathcal{T}, j \in \mathcal{Z}}cost^{cl}_{ij} + cost^{chil}_{ij}
\end{split}
\end{equation}

\subsection{Formal Attack Model}
Attack modeling is the process of potential threat recognition and security measures to protect a valuable system. 
\subsubsection{Attacker's Knowledge}
The attacker's knowledge about the system determines the attack's stealthiness. This work considers a knowledgeable attacker aware of the surrounding weather pattern, building attributes, occupancy information, underlying control, and defense mechanisms of the smart building HVAC control system. 

\subsubsection{Attack Goal}
The BIoTA framework considers two different attack goals for launching attacks - increasing energy consumption and obstructing occupants' comfort. The energy consumption attack is straightforward in which the attacker attempts to change the sensor measurements to maximize the overall energy consumption in the building by forcing the control system to flow more fresh air in the supply airflow. 
\begin{equation}
\begin{split}
    \forall_{i \in \mathcal{T}, j \in \mathcal{Z}, k \in [o, ti, ci]} \mathcal{S}^k_{ij} = (\mathcal{S}^k_{ij} + \delta^k_{ij})  \land TC^{HVAC} > \mathbb{HC}
\end{split}
\end{equation}
%
Here, $\delta^k_{ij}$ is the denotes of injected attack for k-th parameter in j-th zone at i-th time instance, while $\mathbb{HC}$ signifies the minimum threshold for cost increase due to attack. The occupants' discomfort attack attempts to deviate the $CO_2$ or temperature of the building zones far away from the setpoint. 
\begin{equation}
\begin{split}
\forall_{i \in \mathcal{T},  j \in \mathcal{Z}, k \in [o, ti, ci]} \mathcal{S}^k_{ij} = (\mathcal{S}^k_{ij} + \delta^k_{ij}) \land \\
 (|{S}_{ci}^{ij} - C^{set}_{ij}| > \mathbb{OCC}_j) \land (|\mathcal{S}_{ti}^{ij} - T^{set}_{ij}| > \mathbb{OCT}_j)
\end{split}
\end{equation}

Here, $\mathbb{OCC}_j$ signifies the minimum threshold for deviation from $CO_2$ concentration setpoint and $\mathbb{OCT}_j$ denotes the deviation of temperature setpoint due to attack.
\subsubsection{Accessibility to Sensor Measurements}
For compromising sensor measurement of a particular zone, the attacker needs to alter the sensor measurements of other sensors of that zone bypassing the controller's anomaly detector. The altered sensor measurements need to comply with controller verification.

\subsubsection{Attacker's Capability}
For successfully launching an attack, the attacker needs to compromise multiple sensor measurements. The altered occupancy sensor measurement of a zone must be less than the maximum capacity.
\begin{equation}
\begin{split}
     \forall_{i \in \mathcal{T}, j \in \mathcal{Z}} \mathcal{S}^{o}_{ij} + \delta_{ij} \leq \mathbb{MO}_j
\end{split}
\end{equation}
Here, $\mathbb{MO}_j$ denotes the maximum occupancy at the j-th zone.
The attackers are constrained to change the occupancy sensor measurements without making alternations in the total actual occupant count of the current timeslots. Again, other sensor needs to be altered accordingly to avoid getting exposed.
\begin{equation}
\begin{split}
     \forall_{i \in \mathcal{T}} \sum_{j \in \mathcal{Z}}(\mathcal{S}^{o}_{ij} + \delta_{ij}) = \sum_{j \in \mathcal{Z}}(\mathcal{S}^{o}_{ij}) 
\end{split}
\end{equation}


\section{Example Case Study}
\label{sec:case-study}

We perform two case studies on both commercial and residential datasets. 
Table~\ref{tab:testbed-description} shows the properties of the zones of the datasets.

\subsection{Case Study on the COD Dataset}
To verify our BIoTA framework, we collect occupancy data from the COD dataset, which captures almost 90 thousand people entering and leaving five different spaces~\cite{liu2017cod}.
We analyze one year of data with ten-minute sampling by generating missing data from the data distribution. Table~\ref{tab:case_study} shows actual sensor measurements and required average ventilation rate for the zones ~\cite{climate2020}.

\vspace{3pt}
\noindent\textbf{Energy Consumption Attack:} Suppose an attacker is injecting false sensor measurements to launch a stealthy attack with the intent to increase the energy consumption of the building by 5\%. Table~\ref{tab:case_study} demonstrates a possible attack scenario, where the attacker achieves his attack goal by adding and subtracting occupancy sensor measurements by 17 to the entrance and the Warhol zones, respectively.
At entrances, people produce 12\% more $CO_2$ and radiates 17\% more heat than the Warhol zones. Hence, due to the compromised measurements, the controller supplies 56\% more air than required in the entrances, where the Warhol zone gets 13 times less airflow. Eventually, the entrance zone's temperature increases by 4\%, whereas in Warhol zones, the temperature falls near the outdoor temperature. The actual energy requirement is $79.69$ kWh for the presented scenario, but for compromised sensor measurements, the energy consumption goes higher to $82.23$ kWh. The cost of electricity of Pennsylvania for wholesale electricity market is \$ $7.02$  per kWh based on the recent rates~\cite{electricrate2020}. So, there is a $3$\textcent of overall cost increase in $10$ minutes time interval because of the increased energy requirement. The attack analysis also reports that the same attack goal can be achieved by altering Warhol and laboratory measurements only, which in turn increases the energy consumption by $3.47$ kWh and cost by $4.06$\textcent.

\vspace{3pt}
\noindent\textbf{Occupant Comfort Attack} For performing occupants' comfort attack, let's assume, the attacker is modifying a set of sensor measurements to launch a stealthy attack with the intent to increase the $CO_2$ concentration of the laboratory zone by 25\%. Table~\ref{tab:case_study} demonstrates a possible attack scenario, where the attacker achieves his attack goal by compromising occupancy sensors of the entrance and the laboratory zones by altering occupancy measurements by 20 people from Warhol to the entrance zone. In the laboratory zone, people produce 15\% more $CO_2$ and radiates 10\% more heat than the entrances. Hence, the controller supplies 66\% more air than required in the entrances, where the laboratory zone will get four times less airflow. Eventually, the temperature of entrance zones increases by 5\%, and in laboratory zones, the temperature falls near the outdoor temperature. 
For the presented scenario, the actual energy requirement is reduced by $1.69$ kWh with the cost of significant discomfort to the occupants of the laboratory, which hampers the industry's productivity. The attack analysis shows that the same attack goal cannot be achieved by modifying any other two sensors.

\begin{table*}[t]
\centering
\caption{Example Stealthy Attack Scenarios in COD and KTH datasets}
\label{tab:case_study}
\scriptsize
\vspace{-6pt}
\begin{tabular}{|c|l|c|c|c|c|c|c|c|}
\hline
\multirow{2}{*}{\textbf{Scenario}}                                                              & \multicolumn{1}{c|}{\multirow{2}{*}{\textbf{Measurement}}} & \multicolumn{4}{c|}{\textbf{COD Dataset}}                                                                                                                  & \multicolumn{3}{c|}{\textbf{KTH Dataset}}                                                                                  \\ \cline{3-9} 
                                                                                                & \multicolumn{1}{c|}{}                                      & \multicolumn{1}{l|}{\textbf{Entrance}} & \multicolumn{1}{l|}{\textbf{Clemente}} & \multicolumn{1}{l|}{\textbf{Warhol}} & \multicolumn{1}{l|}{\textbf{Lab}} & \multicolumn{1}{l|}{\textbf{Living Room}} & \multicolumn{1}{l|}{\textbf{Kitchen}} & \multicolumn{1}{l|}{\textbf{Bathroom}} \\ \hline
\multirow{4}{*}{\textbf{Actual}}                                                                & \textbf{Zone Occupancy (Person)}                           & 30                                     & 3                                      & 18                                   & 26                                & 4                                         & 1                                     & 0                                      \\ \cline{2-9} 
                                                                                                & \textbf{Zone Temperature (degree C)}                       & 75                                     & 75                                     & 75                                   & 75                                & 69.8                                      & 68.9                                  & 69.5                                   \\ \cline{2-9} 
                                                                                                & \textbf{Zone CO2 (PPM)}                                    & 1000                                   & 1000                                   & 1000                                 & 1000                              & 927.1                                     & 923.5                                 & 854.7                                  \\ \cline{2-9} 
                                                                                                & \textbf{Zone Airflow (CFM)}                            & 1105.0                                 & 106.5                                  & 549.0                                & 1100.66                           & 138.87                                    & 42.21                                 & 1.3                                    \\ \hline
\multirow{4}{*}{\textbf{\begin{tabular}[c]{@{}c@{}}Energy\\ Consumption\\ Attack\end{tabular}}} & \textbf{Zone Occupancy (Person)}                           & 47                                     & 3                                      & 1                                    & 26                                & 2                                         & 2                                     & 1                                      \\ \cline{2-9} 
                                                                                                & \textbf{Zone CO2 (PPM)}                                    & 704.26                                 & 1000                                   & 1282.13                              & 1000                              & 1254                                      & 400                                   & 400                                    \\ \cline{2-9} 
                                                                                                & \textbf{Zone Temperature (degree C)}                       & 78.67                                  & 75                                     & 62                                   & 75                                & 60.9                                      & 73.1                                  & 76.6                                   \\ \cline{2-9} 
                                                                                                & \textbf{Zone Airflow (CFM)}                            & 1724                                   & 106.5                                  & 41.09                                & 1100.66                           & 69.43                                     & 84.43                                 & 55.86                                  \\ \hline
\multirow{4}{*}{\textbf{\begin{tabular}[c]{@{}c@{}}Occupant\\ Comfort \\ Attack\end{tabular}}}  & \textbf{Zone Occupancy (Person)}                           & 50                                     & 3                                      & 18                                   & 6                                 & 5                                         & 0                                     & 0                                      \\ \cline{2-9} 
                                                                                                & \textbf{Zone CO2 (PPM)}                                    & 648.38                                 & 1000                                   & 1000                                 & 1283.2                            & 763.59                                    & 156.45                                & 854.7                                  \\ \cline{2-9} 
                                                                                                & \textbf{Zone Temperature (degree C)}                       & 79.08                                  & 75                                     & 62                                   & 62                                & 71.5                                      & 39                                    & 69.5                                   \\ \cline{2-9} 
                                                                                                & \textbf{Zone Airflow (CFM)}                            & 1841.66                                & 106.5                                  & 549.0                                & 254.0                             & 173.6                                     & 1.7                                 & 1.3                                    \\ \hline
\end{tabular}
\vspace{-12pt}
\end{table*}

\subsection{Case Study on KTH Live-In Lab Dataset}
We also verify our BIoTA framework using the KTH Live-In Lab dataset to assess the vulnerability of residential buildings. We analyze one-week data with ten-minute sampling by generating missing data from the available data distribution. We consider each room as a zone and the apartment to be a building for making HVAC control decisions. Table~\ref{tab:case_study} shows actual sensor measurements and the required average airflow rate for the three zones. 
\vspace{3pt}

\noindent\textbf{Energy Consumption Attack:} Suppose an attacker intends to increase the energy consumption of the considered residential building by more than 10\%. Table~\ref{tab:case_study} demonstrates a possible attack scenario, where the attacker achieves his attack goal compromising occupancy sensors of all three zones by adding one occupant to the bathroom and kitchen occupancy sensor measurements and balance it by removing 2 person from the living room occupancy sensor measurement. Therefore, the controller supplies 35\% more air than required in the living room, where in the bathroom and the kitchen zones, the $CO_2$ concentration gets lower down to the outdoor air concentration. Eventually, the temperature of the living room zone falls by 13\%, where in the kitchen and bathroom, the temperature rises by 6\% and 10\% respectively. For the presented scenario, the actual energy requirement is 16.41 kWh, but for compromised sensor measurements, the energy consumption increases to 18.37 kWh. The cost of electricity of Stockholm, Sweden for residential electricity market is 24.6 \textcent per kWh based on the 2019 rates~\cite{swedenrate2020}. There will be $3.48$ cents of overall cost increase in 10 minutes of time because of the increased energy requirement. It can be seen that energy consumption attack is also affecting the occupant's comfort as well. BIoTA framework suggests that the attack goal of increasing energy consumption by 10\% cannot be carried out attacking sensor measurements of two zones.

\vspace{3pt}
\noindent\textbf{Occupant Comfort Attack} 
To demonstrate the occupant comfort attack, let's assume that the attacker's goal is to increase the $CO_2$ concentration of the kitchen by more than 25\%. Table~\ref{tab:case_study} shows a possible attack scenario, where the attacker achieves his/her attack intent by compromising the occupancy sensor measurements of the kitchen and living room zone. Due to the attack, the controller supplies 25\% more air than required in the living room. Due to the attack, the living room temperature increases by 2\%, and in kitchen zones, the temperature falls near the outdoor temperature. Hence, the actual energy requirement is reduced by 2.2\% with the cost of significant discomfort to the kitchen occupants. Compromising occupancy sensor measurements of the kitchen and the bathroom can not achieve the attack goal as suggested by our framework.

\begin{table}[!b]
\centering
\vspace{-15pt}
\caption{95\% of confidence interval for zone-wise occupancy}
\label{tab:stat-data}
\vspace{-6pt}
\scriptsize
\begin{tabular}{|l|l|c|}
\hline
\multicolumn{1}{|c|}{\multirow{2}{*}{\textbf{Dataset}}} & \multirow{2}{*}{\textbf{Zone}} & \multirow{2}{*}{\textbf{\begin{tabular}[c]{@{}c@{}}95\% Confidence Interval\end{tabular}}} \\
\multicolumn{1}{|c|}{}                                  &                                &                                                                                               \\ \hline
\multirow{4}{*}{\textbf{COD}}                           & Entrances                      & 7.88 - 12.36                                                                                  \\ \cline{2-3} 
                                                        & Clemente                       & 0.21-0.55                                                                                     \\ \cline{2-3} 
                                                        & Warhol                         & 0.31 -1.28                                                                                    \\ \cline{2-3} 
                                                        & Laboratory                     & 6.14 - 9.67                                                                                   \\ \hline
\multirow{3}{*}{\textbf{KTH}}                           & Living Room                    & 1.1-1.49                                                                                      \\ \cline{2-3} 
                                                        & Kitchen                        & 0-0.27                                                                                        \\ \cline{2-3} 
                                                        & Bathroom                       & 0-0.32                                                                                        \\ \hline
\end{tabular}
\end{table}

\section{Evaluation}
\label{sec:evaluation}

We implement our proposed BIoTA framework and evaluate it using two real datasets.

\vspace{-3pt}
\subsection{Implementation} 
\label{subsec:setup_env}\vspace{-3pt}
The implementation source code for the BIoTA framework is available in GitHub~\cite{biota2020}.
We use the SMT solver to optimize a set of constraints~\cite{barrett2018satisfiability}. We encode the conditions using integer and real terms. Based on the satisfiability,  the model produces SAT or UNSAT results. The formal threat modeling has been implemented using Z3 library~\cite{de2008z3}. We have used python API of Z3 SMT solver for both solving and optimizing our modeling constraints. The psypy library is used to calculate and extract necessary psychrometric parameters from temperature and relative humidity~\cite{psypy2020}. The implementation of the library follows the ASHRAE standard. 
\begin{figure}[!t]
\vspace{-12pt}
    \begin{center}
        \subfigure[]
        {
        \label{fig:cod_occupancy_pattern}
            \includegraphics[width=0.46\columnwidth]{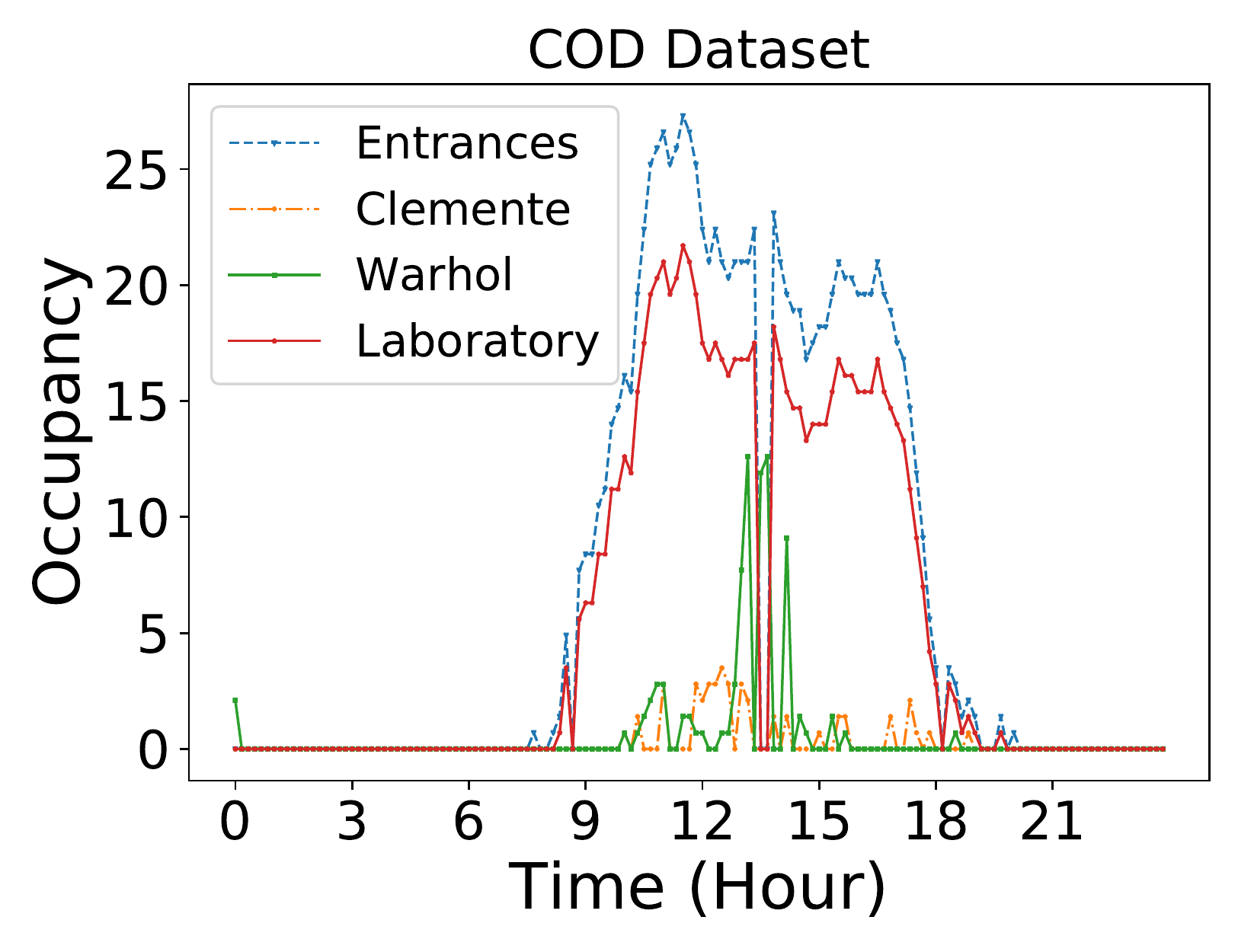}
        }
        \subfigure[]
         {
        \label{fig:kth_occupancy_pattern}
            \includegraphics[width=0.46\columnwidth]{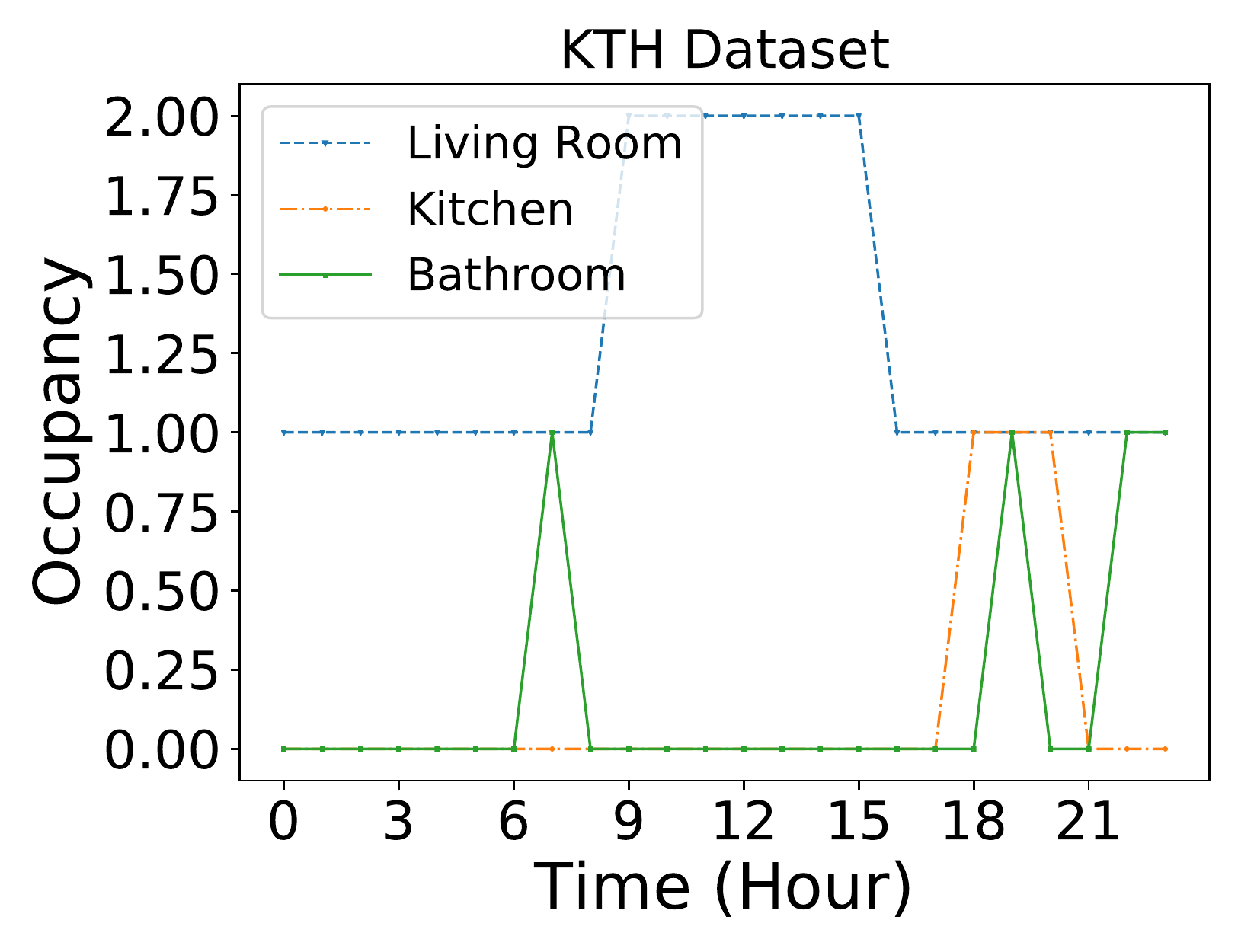}
        }
    \end{center}
    \vspace{-15pt}
    \caption{Occupancy Pattern of (a) COD dataset and (b) KTH dataset.} 
    \vspace{-15pt}
    \label{fig:occupancy pattern}
\end{figure}

\vspace{-3pt}
\subsection{Dataset Description}
\vspace{-3pt}
The occupancy pattern of both COD and KTH datasets are normally distributed. Figs~\ref{fig:cod_occupancy_pattern} and~\ref{fig:kth_occupancy_pattern} shows the average occupancy pattern throughout different hours of the day. The COD dataset captures data for almost a year. For KTH Live-In lab dataset, we only have one-week occupancy and temperature-related data. 
Table~\ref{tab:stat-data} shows the zone-wise mean occupancy information with 95\% confidence interval. In the COD dataset, the laboratory zone is the most occupied, and in the case of the KTH dataset, the living room zone is mostly occupied, probably by some visitors occasionally. The KTH outdoor temperature and relative humidity mean are within (39.3 - 41.2)\textdegree F and (72.3 - 75.9)\% for 95\% confidence interval. In case of COD dataset, temperature mean lies in between (53.3 - 56.9)\textdegree F, and relative humidity (63.39 - 65.96)\% for 95\% confidence interval. We present the source of additional data collection and processing in Section~\ref{sec:case-study}.

\begin{table}[!b]
\centering
\vspace{-15pt}
\caption{Maximum Increase in Energy of COD dataset based on Access to Sensor Measurements of Zones \label{tab:cod_increase_cost}}
\vspace{-6pt}
\scriptsize
\begin{tabular}{|l|c|}
\hline
\textbf{Access to Zones}                                                            & \multicolumn{1}{l|}{\textbf{\begin{tabular}[c]{@{}l@{}}Max Energy Increase (\$)\end{tabular}}} \\ \hline
\begin{tabular}[c]{@{}l@{}}Entrances, Clemente , Warhol, Laboratory\end{tabular} & 303.95                                                                                                 \\ \hline
Entrances, Clemente , Warhol                                                        & 17.14                                                                                                 \\ \hline
Entrances, Warhol, Laboratory                                                       & 245.94                                                                                                 \\ \hline
Entrances, Clemente, Laboratory                                                     & 256.85                                                                                                 \\ \hline

Clemente, Warhol, Laboratory                      
                & 55.24
                                                    \\ \hline
Entrances, Clemente                                                                 & 3.20                                                                                                  \\ \hline
Entrances, Warhol                                                                   & 19.47                                                                                                 \\ \hline
Entrances, Laboratory                                                               & 192.42                                                                                                  \\ \hline
Warhol, Laboratory                                                                  & 45.95                                                                                                 \\ \hline
Clemente, Laboratory                                                                & 8.95                                                                                                  \\ \hline
Clemente , Warhol                                                                   & 3.75                                                                                                 \\ \hline
\end{tabular}
\normalsize
\end{table}

\vspace{-3pt}
\subsection{Evaluation based on Energy Consumption Attack}
\vspace{-3pt}

\begin{figure}[!t]
\centering
    \begin{center}
        \subfigure[]
        {
        \label{fig:energy_consumption_attack_cod}
            \includegraphics[width=0.46\columnwidth]{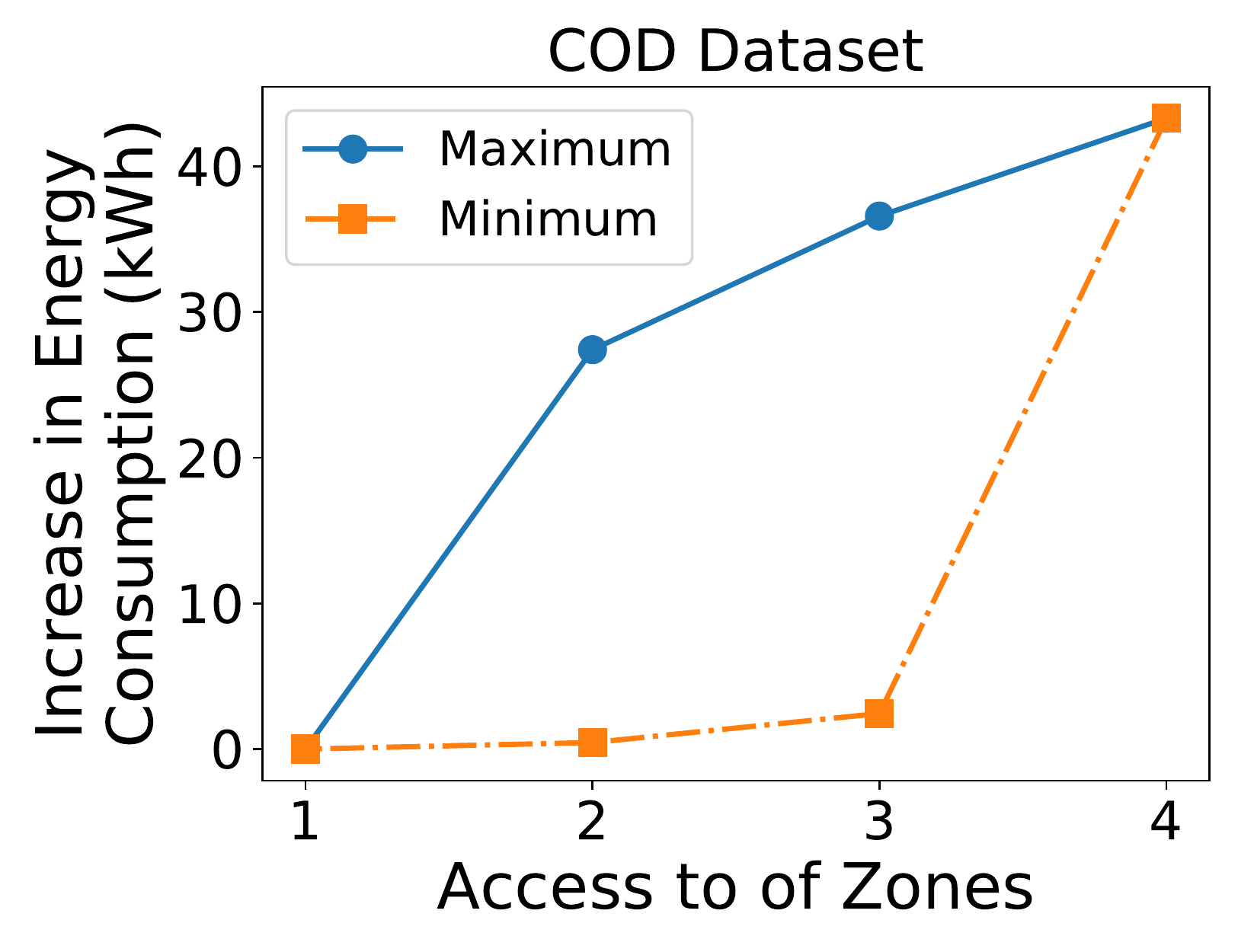}
        }
        \subfigure[]
         {
        \label{fig:energy_consumption_attack_kth}
            \includegraphics[width=0.46\columnwidth]{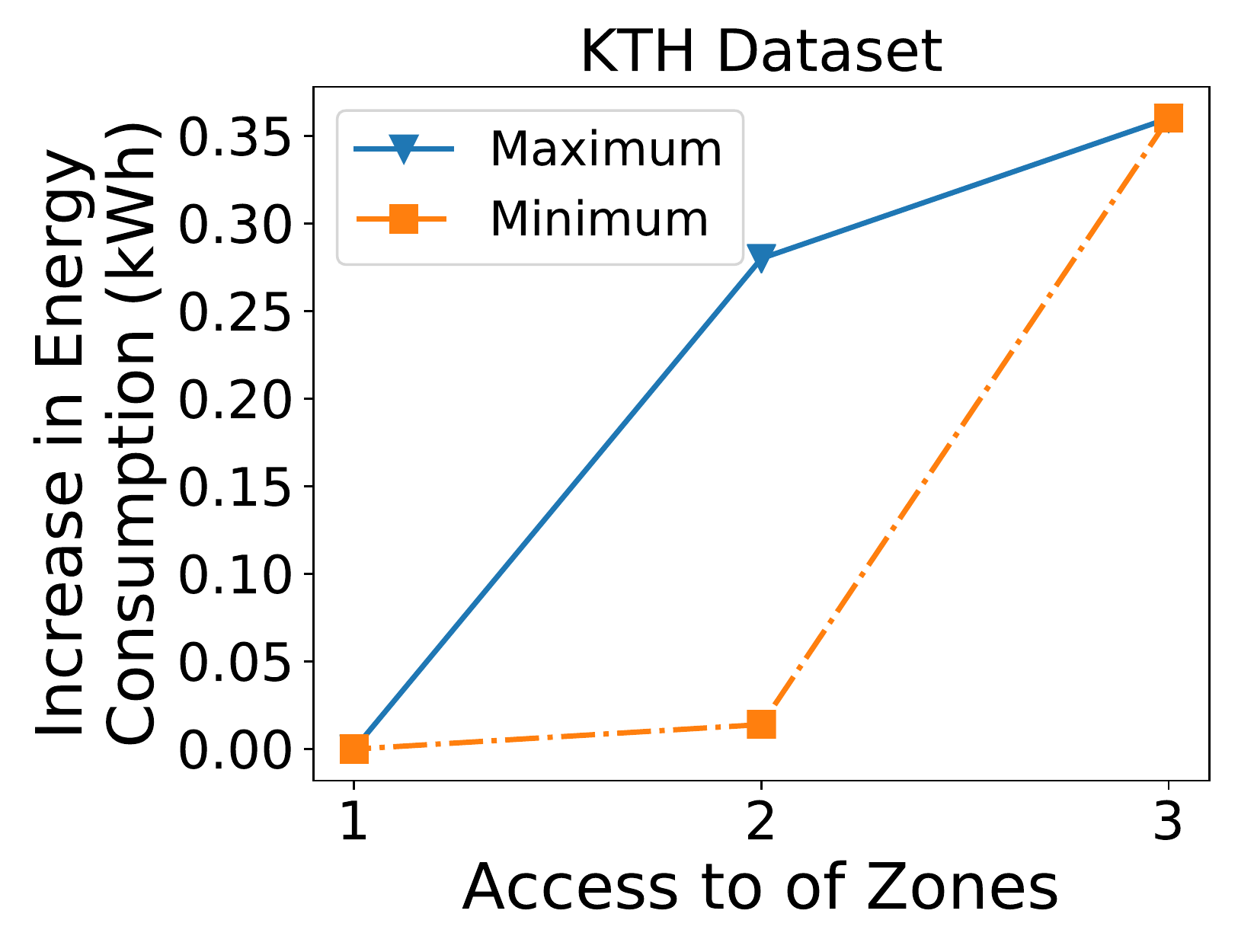}
        }
    \end{center}
    \vspace{-15pt}
    \caption{Increase in the energy consumption: (a) COD and (b) KTH datasets.}
    \vspace{-6pt}
    \label{fig:enegy_consumption_attack}
\end{figure}

\begin{table}[t]
\centering
\caption{Maximum Increase in Energy based of KTH dataset on Access to Sensor Measurements of Zones}
\vspace{-6pt}
\scriptsize
\label{tab:kth_increase_cost}
\begin{tabular}{|l|c|}
\hline
\textbf{Access to Zones}                                                  & \multicolumn{1}{l|}{\textbf{\begin{tabular}[c]{@{}l@{}}Maximum Energy Increase (\$)\end{tabular}}} \\ \hline
\begin{tabular}[c]{@{}l@{}}Living Room, Kitchen, Bathroom\end{tabular} & 8.95                                                                                                   \\ \hline
Living Room, Bathroom                                                     & 7.01                                                                                                  \\ \hline
Living Room, Kitchen                                                      & 2.77                                                                                                   \\ \hline
Kitchen, Bathroom                                                         & 0.35                                                                                                   \\ \hline
\end{tabular}
\normalsize
\vspace{-15pt}
\end{table}

\begin{figure}[t]
    \begin{center}
        \subfigure[]
        {
        \label{fig:cod_co2_discomfort}
            \includegraphics[width=0.46\columnwidth]{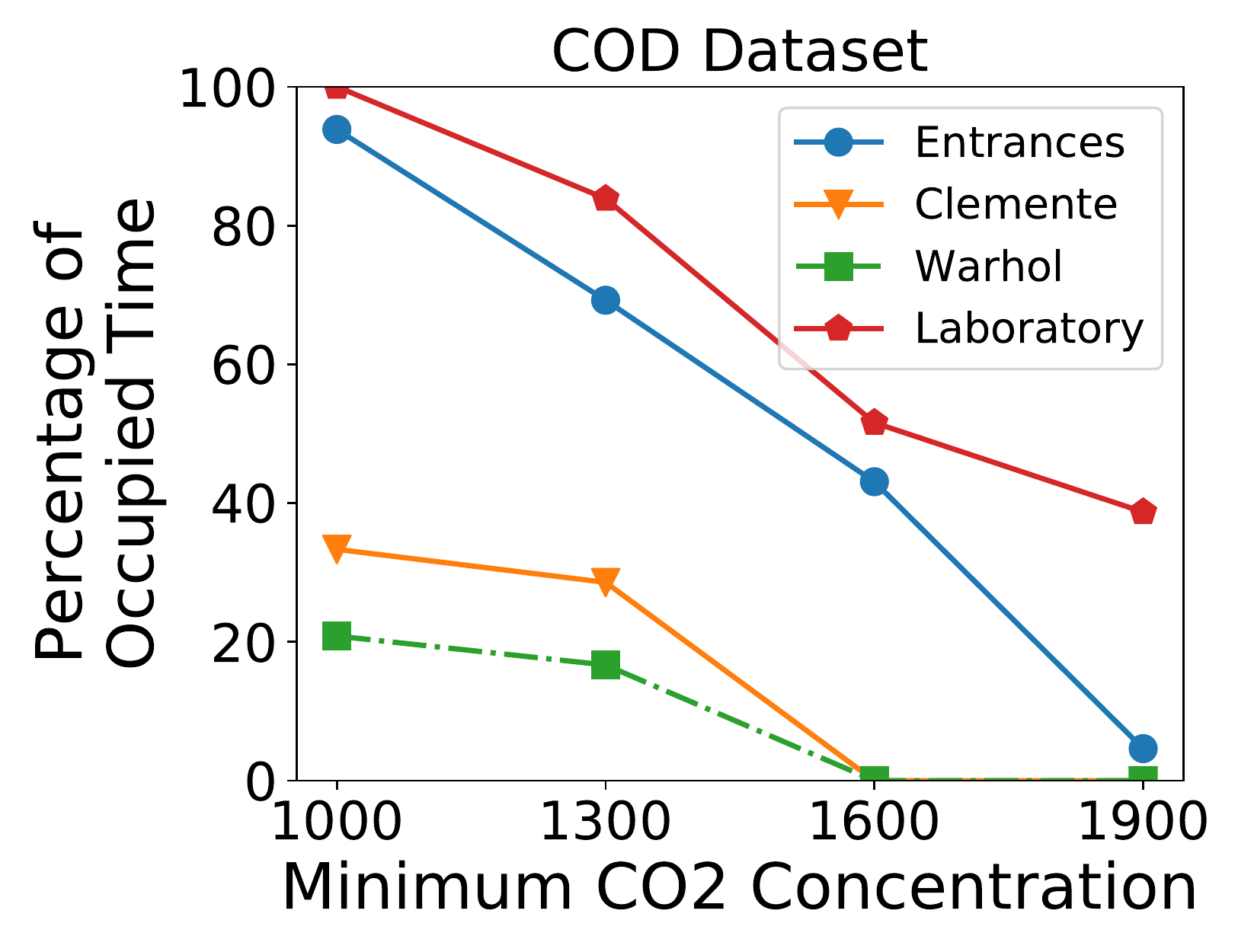}
        }
        \subfigure[]
         {
        \label{fig:kth_co2_discomfort}
            \includegraphics[width=0.46\columnwidth]{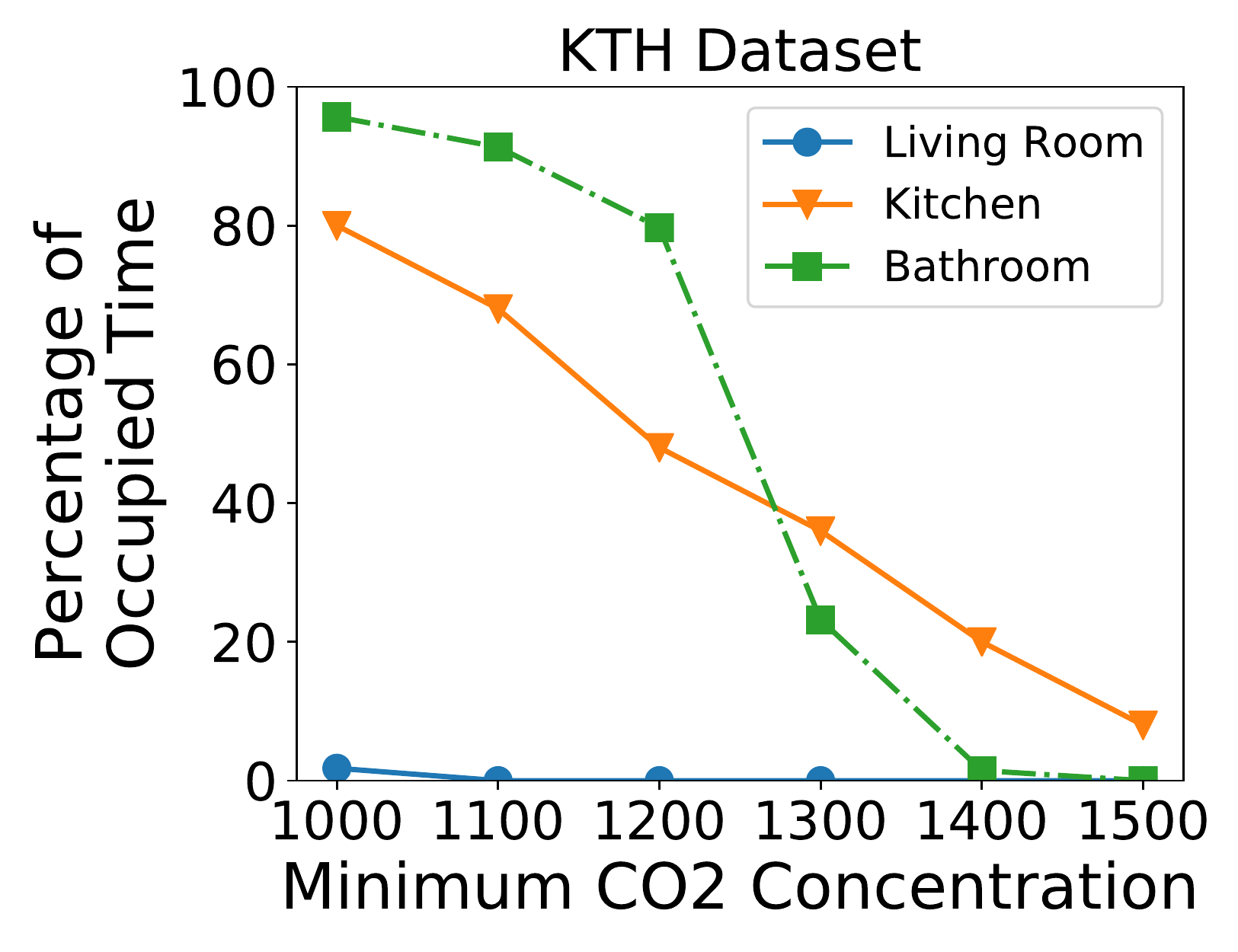}
        }
    \end{center}
    \vspace{-15pt}
    \caption{Occupants' discomfort measure based on $CO_2$ concentration: (a) COD dataset and (b) KTH dataset.}
    \vspace{-6pt}
    \label{fig:co2_discomfort}
\end{figure}

\begin{figure}[t]
    \begin{center}
        \subfigure[]
        {
        \label{fig:cod_temp_discomfort}
            \includegraphics[width=0.46\columnwidth]{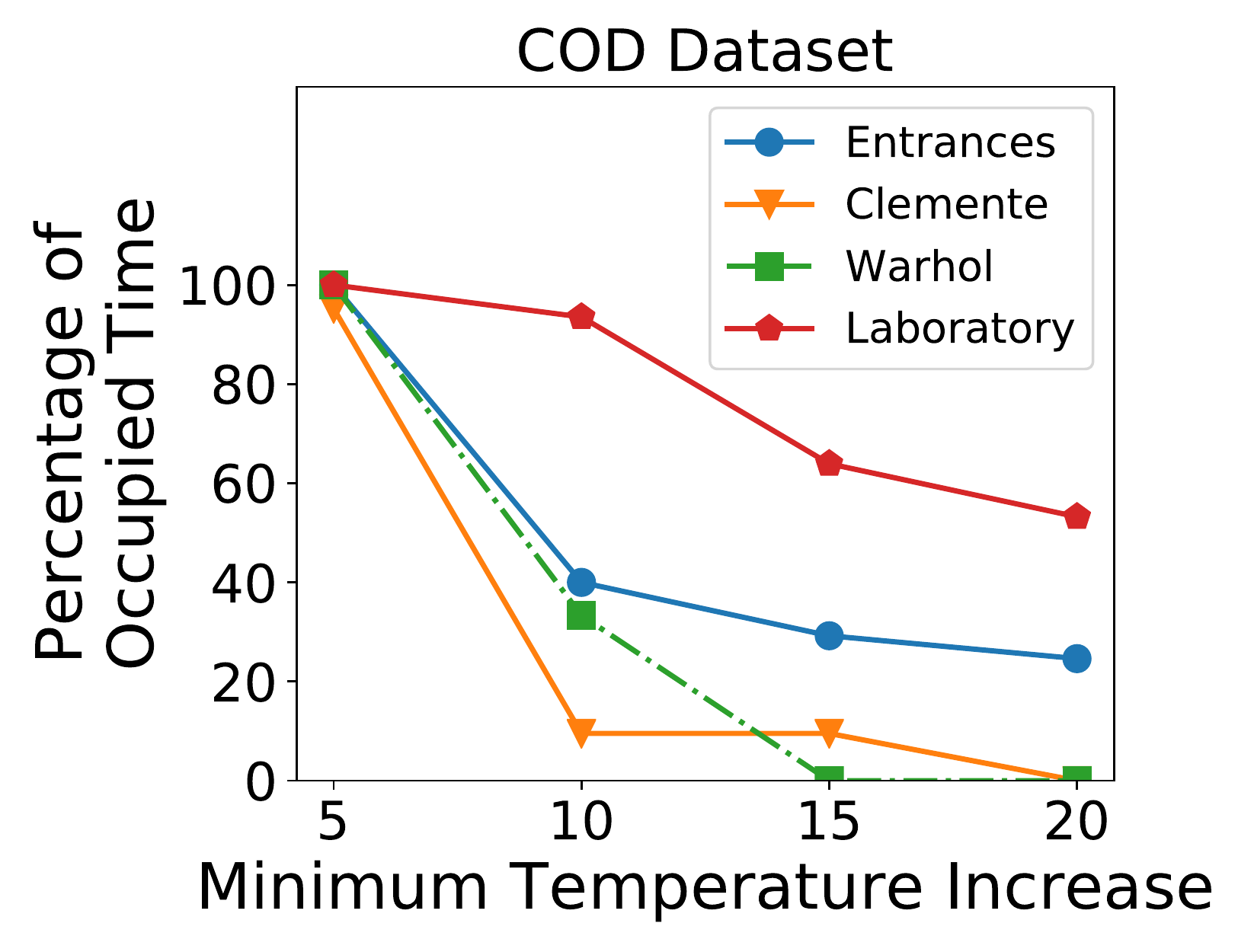}
        }
        \subfigure[]
         {
        \label{fig:kth_temp_discomfort}
            \includegraphics[width=0.46\columnwidth]{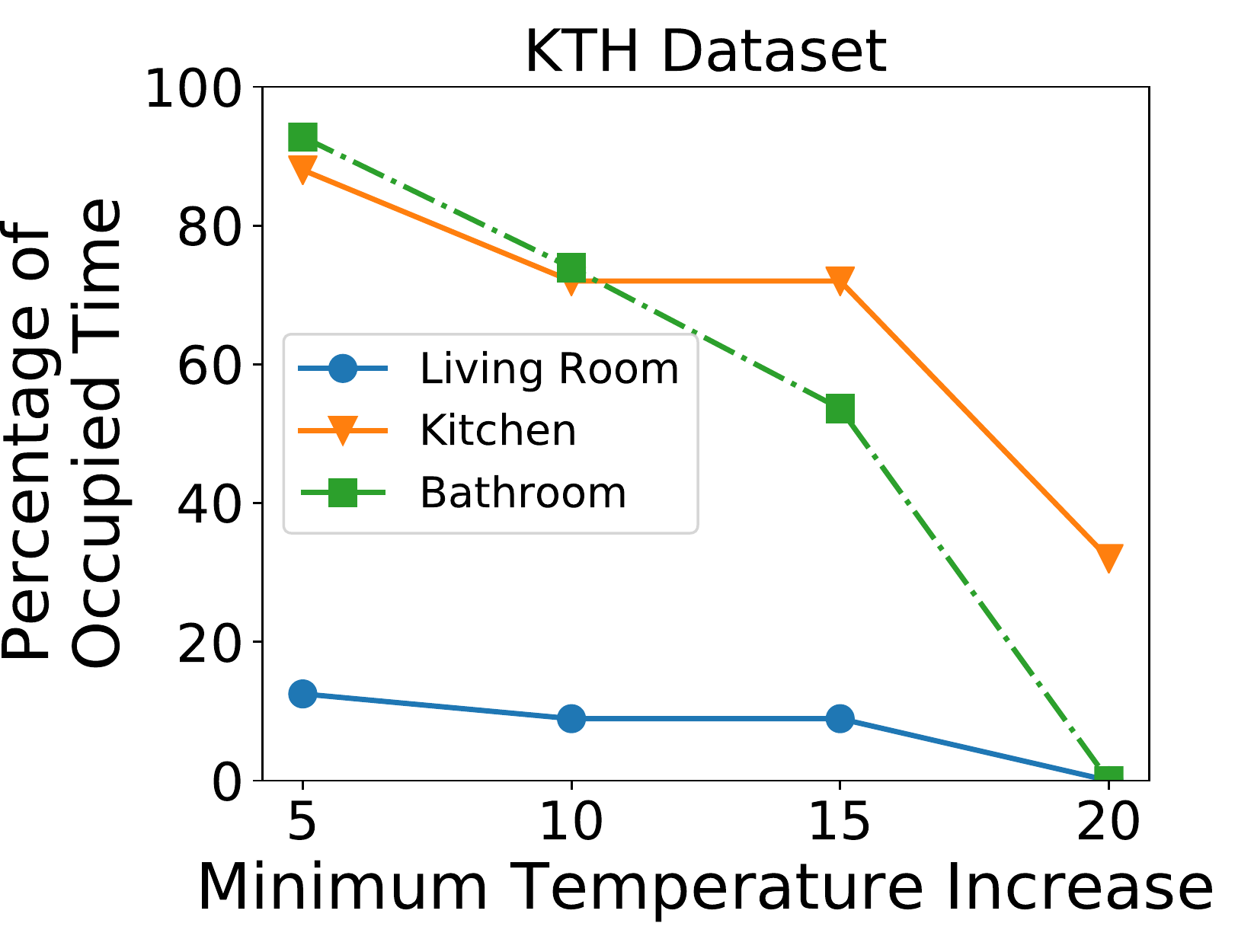}
        }
    \end{center}
    \vspace{-15pt}
    \caption{Occupants' discomfort measure based on temperature: (a) COD dataset and (b) KTH dataset.}
    \vspace{-12pt}
    \label{fig:temp_discomfort}
\end{figure}

To assess the maximum attack impact, we consider that an attacker is attacking every timeslot with optimal attack targets for maximizing the overall energy cost of the building. Fig.~\ref{fig:energy_consumption_attack_cod} shows the maximum and minimum attack impact based on energy consumption for a varying number of zone sensor measurements access in the COD dataset when the attacker is attacking optimally on all the timeslots. Table~\ref{tab:cod_increase_cost} shows the maximum increase in cost for all combination of zone sensor measurement access. It can be depicted that an attacker cannot do much harm if he/she has no access to sensor measurements of the laboratory zone. As people in the laboratory tend to work harder than the other zones, the controller needs to flow more air per occupant for this zone. Hence, the optimal attack goal will try to alter measurements showing occupants' movement from other zones to laboratory zones. As in Fig.~\ref{fig:energy_consumption_attack_cod}, there is no increase in energy cost for only one zone sensor measurement access as changing the occupancy of a single zone will eventually fail the verification of the total number of people at a particular control instance.  Table~\ref{tab:kth_increase_cost} shows the maximum increase in cost for all combinations of zone sensor measurement access. 

Fig.~\ref{fig:energy_consumption_attack_kth} shows the attack impact based on energy consumption for a varying number of zone sensor measurement access, and Table~\ref{tab:kth_increase_cost} shows the maximum increase in cost for all combination of zone sensor measurement access for the KTH Live-In lab dataset. If an attacker can access two zone's sensor measurements, maximum attack impact can be attained if he/has access to all sensor measurements of the living room and bathroom. This is because the living room zone has most people to be swapped in the two other rooms. 

\begin{figure}[t]
    \begin{center}
        \subfigure[]
        {
        \label{fig:cod_scalability}
            \includegraphics[width=0.46\columnwidth]{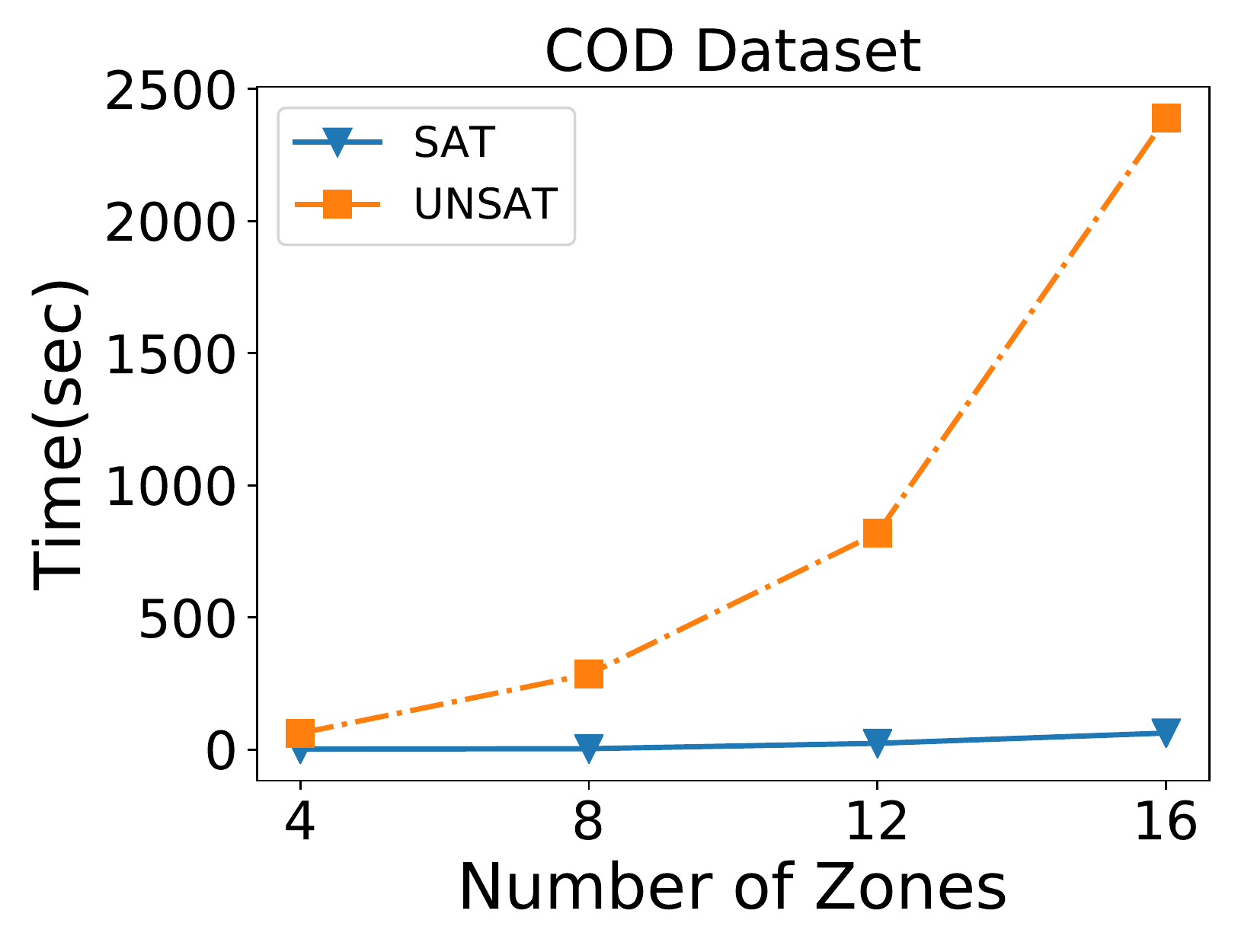}
        }
        \subfigure[]
         {
        \label{fig:kth_scalability}
            \includegraphics[width=0.46\columnwidth]{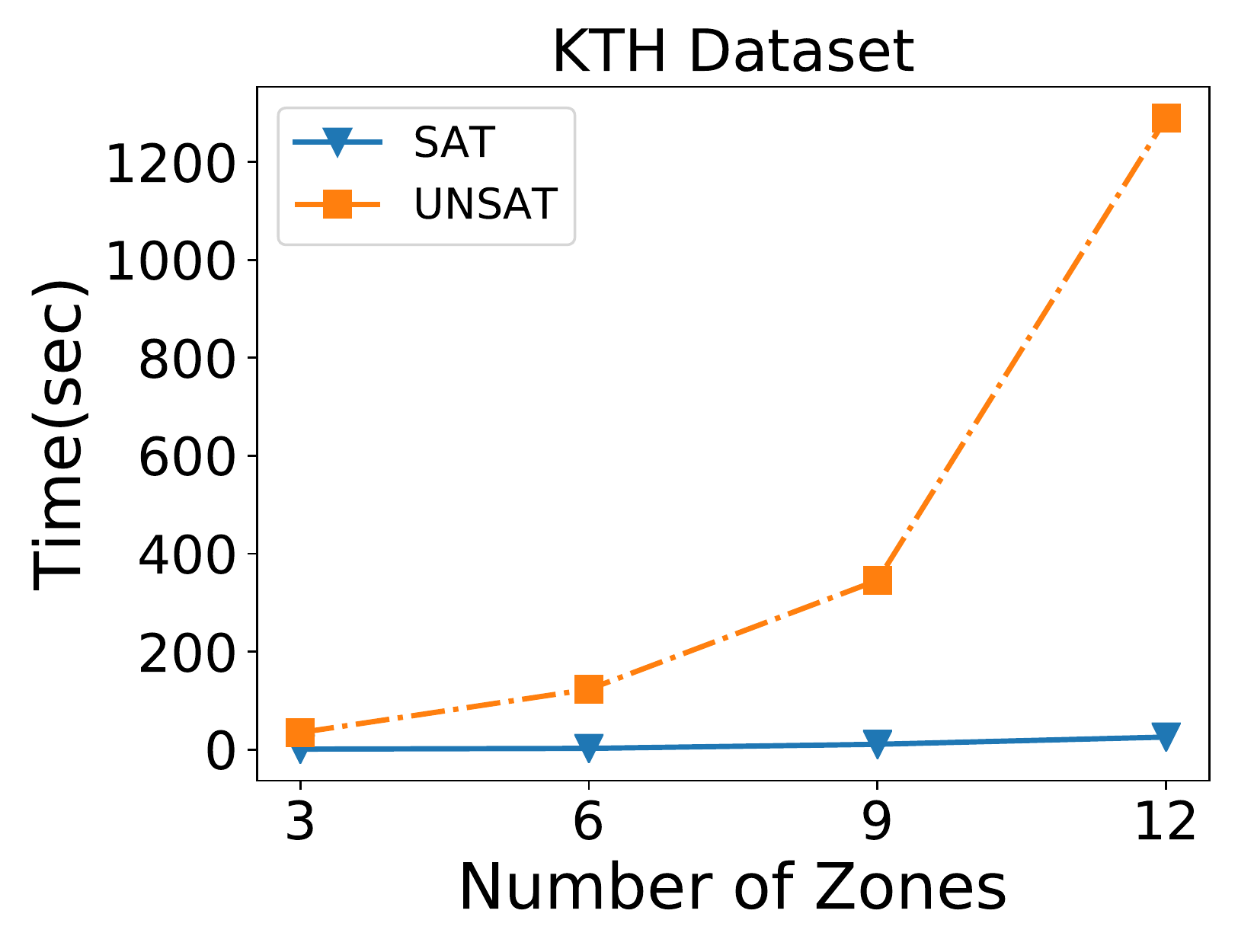}
        }
    \end{center}
    \vspace{-15pt}
    \caption{Scalability analysis: (a) COD dataset and (b) KTH dataset.}
    \vspace{-6pt}
    \label{fig:scalability}
\end{figure}

\vspace{-3pt}
\subsection{Evaluation based on Occupants' Comfort Attack}
\vspace{-3pt}

Our proposed framework performs a two-step evaluation of attacks on occupants' comfort.
At first, the attack impact is analyzed, considering the attack goal is to disrupt the occupants' ventilation need. Depending on the zones' sensor accessibility, we assume that the attacker is conducting optimal attacks in every timestep to ensure maximum discomfort in the maximum zones. Fig.~\ref{fig:cod_co2_discomfort} and~\ref{fig:kth_co2_discomfort} demonstrates the percentage of time $CO_2$ concentration of various zones of our dataset in consideration were failed to maintain IAQ due to stealthy FDI attack. It can be seen that if an attacker has access to all sensor measurements of the zones, attack analysis of the COD dataset says that almost 4\% of the occupied time, $CO_2$ concentration in entrance zone exceeds 1900 ppm throughout the year while for laboratory zone it is almost 38\%. Analysis of the KTH dataset indicates that 8\% of the occupied time, the kitchen's $CO_2$ concentration exceeds 1500ppm. 

Attack analysis on occupants' comfort attack shows that if an attacker has access to all sensor measurements of the zones, 
the temperature of the zones in entrance zone exceeds more than 20 \textdegree F almost 24.6\% of the occupied time throughout the year while for laboratory zone it is almost 53.22\%. Analysis of the KTH dataset indicates that during 32\% of the occupied time, kitchen zone temperature exceeds more than 20 \textdegree F throughout the week. For the bathroom, it is almost 53.62\% of the occupied time. 

\vspace{-3pt}
\subsection{Scalability}
\vspace{-3pt}
We evaluate the scalability of BIoTa framework by varying the sizes of the zones of the available dataset. The increase in the number of zones multiplies the number of constraints. 
The occupancy of the new zones follows the distribution of the available zones. The SMT solver takes significantly less time for SAT cases, as seen from Fig.~\ref{fig:scalability}. 
For the COD dataset, the SAT cases take 1.59, 3.14,  23.42, and 62.3 seconds respectively for 4, 8, 12, and 16 zones respectively, where for the UNSAT case it took 61.8, 285.1, 819.2, and 2389.2 seconds.  For the KTH dataset, SAT cases take 1.02, 2.5,  10.82, and 25.6 seconds respectively for 3, 6, 9, and 12 zones respectively, where for the UNSAT case it took 34.8, 123.1, 345.7, and 1289.2 seconds. The addition of zones affects the solver's solving time almost exponentially, which proves that the solver is not scalable.

\begin{table}[]
\centering
\caption{Complexity Analysis of BIoTA} 
\label{tab:complexity}
\vspace{-6pt}
\scriptsize
\begin{tabular}{|l|l|c|c|}
\hline
\multirow{2}{*}{\textbf{Dataset}} & \multicolumn{3}{l|}{\textbf{Number of Clauses}}                                                                 \\ \cline{2-4} 
                                  & \textbf{Control}        & \multicolumn{1}{l|}{\textbf{Cost Calculation}} & \multicolumn{1}{l|}{\textbf{Attack}} \\ \hline
\textbf{COD}                      & \multicolumn{1}{c|}{52} & 28                                             & 16                                   \\ \hline
\textbf{KTH}                      & \multicolumn{1}{c|}{39} & 21                                             & 12                                   \\ \hline
\end{tabular}
\vspace{-15pt}
\normalsize
\end{table}

The scalability depends on the number of clauses. We analyze the complexity of our solver based on the number of clauses. Table~\ref{tab:complexity} shows the number of clauses for a single zone. The addition of new zones adds an almost similar number of new clauses to the solver, which enlarges the solution domain significantly.


\vspace{-3pt}
\section{Conclusion}
\label{sec:conclusion}\vspace{-3pt}

The BIoT-based HVAC control system is gaining popularity because of its reliability, efficacy, and cost-efficiency. But our attack analysis on the smart building  HVAC control system using the BIoTA framework suggests that exploitation of vulnerable sensor measurements of smart building HVAC control system can result in a significant increase in cost and raises health concerns of the occupants. Our proposed attack model is verified using two state-of-the-art building occupancy datasets. 
The COD dataset's evaluation results show that using an energy consumption attack, the HVAC control system can encounter an additional cost of \$300.
We plan to study cyber attacks that exploit multiple smart building control systems, running independently, to create a deeper, cascaded impact.

\bibliographystyle{unsrt}
\bibliography{References}

\begin{thebibliography}{10}

\bibitem{encycle2020}
Encycle delivers instant energy savings with swarm logic®, our patented
  iot-enabled technology.
\newblock \url{https://www.encycle.com/swarm-logic/}, 2020.
\newblock Accessed: 2020-10-21.

\bibitem{bajer2018iot}
Marcin Bajer.
\newblock Iot for smart buildings-long awaited revolution or lean evolution.
\newblock In {\em IEEE 6th International Conference on Future Internet of
  Things and Cloud (FiCloud)}, pages 149--154, 2018.

\bibitem{akkaya2015iot}
K.~Akkaya, I.~Guvenc, R.~Aygun, N.~Pala, and A.~Kadri.
\newblock Iot-based occupancy monitoring techniques for energy-efficient smart
  buildings.
\newblock In {\em IEEE Wireless Communications and Networking Conference
  Workshops}, pages 58--63, 2015.

\bibitem{yu2010integrating}
Yuebin Yu.
\newblock Integrating air handling units in office buildings for high
  performance.
\newblock 2010.

\bibitem{target2014}
Target attack shows danger of remotely accessible hvac systems.
\newblock
  \url{https://www.computerworld.com/article/2487452/target-attack-shows-danger-of-remotely-accessible-hvac-systems.html},
  2014.
\newblock Accessed: 2020-10-13.

\bibitem{asmag2019}
Smart hvac systems vulnerable to being controlled by hackers through cyber
  attacks.
\newblock \url{https://www.asmag.com/showpost/30716.aspx}, 2019.
\newblock Accessed: 2020-10-13.

\bibitem{IBTimes}
International Business~Times India~Ashok.
\newblock Hackers leave finnish residents cold after ddos attack knocks out
  heating systems.
\newblock
  {http://www.ibtimes.co.uk/hackers-leave-finnish-residents-cold-after-ddos-attack-knocks-out-heating-systems-1590639},
  November 2016.
\newblock Accessed: 2020-07-20.

\bibitem{lu2011novel}
Tao Lu, Xiaoshu L{\"u}, and Martti Viljanen.
\newblock A novel and dynamic demand-controlled ventilation strategy for co2
  control and energy saving in buildings.
\newblock {\em Energy and buildings}, 43(9):2499--2508, 2011.

\bibitem{liu2017cod}
K.~Liu, E.~Pinto, S.~Munir, J.~Francis, C.~Shelton, M.~Berges, and S.~Lin.
\newblock Cod: a dataset of commercial building occupancy traces.
\newblock In {\em ACM International Conference on Systems for Energy-Efficient
  Built Environments}, pages 1--2, 2017.

\bibitem{kth2020}
Testbed kth sample data, gdpr compliance.
\newblock
  \url{https://www.liveinlab.kth.se/en/projekt/2.90711/download-sample-datasheets-1.974705}.
\newblock Accessed: 2020-05-21.

\bibitem{leong2019fault}
Cheng~Yew Leong.
\newblock Fault detection and diagnosis of air handling unit: A review.
\newblock In {\em MATEC Web of Conferences}, volume 255, page 06001. EDP
  Sciences, 2019.

\bibitem{wang2012air}
Gang Wang and Li~Song.
\newblock Air handling unit supply air temperature optimal control during
  economizer cycles.
\newblock {\em Energy and Buildings}, 49:310--316, 2012.

\bibitem{inspectapedia2020}
Warm air heat temperatures \& supply air temperature improvement.
\newblock \url{https://inspectapedia.com/heat/Warm_Air_Supply.php/}.
\newblock Accessed: 2020-06-21.

\bibitem{Ohsonline2016}
An assessment of energy technologies and research opportunities.
\newblock
  \url{https://ohsonline.com/Articles/2016/04/01/Carbon-Dioxide-Detection-and-Indoor-Air-Quality-Control.aspx?Page=2},
  2015.
\newblock Accessed: 2020-10-13.

\bibitem{cali2015co2}
D.~Cali, P.~Matthes, K.~Huchtemann, R.~Streblow, and D.~M{\"u}ller.
\newblock Co2 based occupancy detection algorithm: Experimental analysis and
  validation for office and residential buildings.
\newblock {\em Building and Environment}, 86:39--49, 2015.

\bibitem{dyro2004clinical}
Joseph Dyro.
\newblock {\em Clinical engineering handbook}.
\newblock Elsevier, 2004.

\bibitem{persily2017carbon}
Andrew Persily and Lilian de~Jonge.
\newblock Carbon dioxide generation rates for building occupants.
\newblock {\em Indoor air}, 27(5):868--879, 2017.

\bibitem{tu2019trick}
Y.~Tu, S.~Rampazzi, B.~Hao, A.~Rodriguez, K.~Fu, and X.~Hei.
\newblock Trick or heat? manipulating critical temperature-based control
  systems using rectification attacks.
\newblock In {\em ACM SIGSAC Conference on Computer and Communications
  Security}, pages 2301--2315, 2019.

\bibitem{newaz2020adversarial}
AKM Newaz, Nur~Imtiazul Haque, Amit~Kumar Sikder, Mohammad~Ashiqur Rahman, and
  A~Selcuk Uluagac.
\newblock Adversarial attacks to machine learning-based smart healthcare
  systems.
\newblock {\em arXiv preprint arXiv:2010.03671}, 2020.

\bibitem{mace2018multi}
JC~Mace, C~Morisset, K~Pierce, C~Gamble, C~Maple, and J~Fitzgerald.
\newblock A multi-modelling based approach to assessing the security of smart
  buildings.
\newblock 2018.

\bibitem{meyer2016threat}
D.~Meyer, J.~Haase, M.~Eckert, and B.~Klauer.
\newblock A threat-model for building and home automation.
\newblock In {\em 2016 IEEE 14th international conference on industrial
  informatics (INDIN)}, pages 860--866, 2016.

\bibitem{mustafa2016ata}
Q.~Mustafa, A.~Waleed, and V.S. Kharchenko.
\newblock Ata-based security assessment of smart building automation systems.
\newblock {\em Radioelectronic and computer systems}, (3):30--40, 2016.

\bibitem{abdulmunem2016availability}
A.~Abdulmunem and V.~Kharchenko.
\newblock Availability and security assessment of smart building automation
  systems combining of attack tree analysis and markov models.
\newblock In {\em Intl. Conference on Mathematics and Computers in Sciences and
  in Industry (MCSI)}, pages 302--307, 2016.

\bibitem{hachem2020modeling}
J.~Hachem, V.~Chiprianov, M.~Babar, T.~Khalil, and P.~Aniorte.
\newblock Modeling, analyzing and predicting security cascading attacks in
  smart buildings systems-of-systems.
\newblock {\em Journal of Systems and Software}, 162, 2020.

\bibitem{haque2021novel}
Nur~Imtiazul Haque, Mohammad~Ashiqur Rahman, Md~Hasan Shahriar, Alvi~Ataur
  Khalil, and Selcuk Uluagac.
\newblock A novel framework for threat analysis of machine learning-based smart
  healthcare systems.
\newblock {\em arXiv preprint arXiv:2103.03472}, 2021.

\bibitem{climate2020}
The pennsylvania state climatologist.
\newblock \url{http://www.climate.psu.edu/data/city_information/index.php}.
\newblock Accessed: 2020-06-21.

\bibitem{electricrate2020}
Pennsylvania commercial electric rates.
\newblock \url{https://www.electricrate.com/ commercial-rates/pennsylvania/}.
\newblock Accessed: 2020-06-21.

\bibitem{swedenrate2020}
Electricity prices for households in sweden 2010-2019, semi-annually.
\newblock
  \url{https://www.statista.com/statistics/418124/electricity-prices-for-households-in-sweden/}.
\newblock Accessed: 2020-06-21.

\bibitem{biota2020}
Biota.
\newblock
  \url{https://github.com/imtiazulhaque/research-implementations/tree/main/biota},
  2020.

\bibitem{barrett2018satisfiability}
Clark Barrett and Cesare Tinelli.
\newblock Satisfiability modulo theories.
\newblock In {\em Handbook of Model Checking}, pages 305--343. Springer, 2018.

\bibitem{de2008z3}
Leonardo De~Moura and Nikolaj Bj{\o}rner.
\newblock Z3: An efficient smt solver.
\newblock In {\em International conference on Tools and Algorithms for the
  Construction and Analysis of Systems}, pages 337--340. Springer, 2008.

\bibitem{psypy2020}
Psypy project description.
\newblock \url{https://pypi.org/project/psypy/}.
\newblock Accessed: 2020-03-19.

\end{thebibliography}
\end{document}